\pdfoutput=1
\newcommand*{\ATLASLATEXPATH}{latex/}
\documentclass[cernpreprint,texlive=2016,txfonts,UKenglish]{\ATLASLATEXPATH atlasdoc}

\usepackage[subfigure]{\ATLASLATEXPATH atlaspackage}
\usepackage{\ATLASLATEXPATH atlasbiblatex}

\usepackage[process]{\ATLASLATEXPATH atlasphysics}

\graphicspath{{logos/}{figures/}}

\usepackage{paper-defs}

\usepackage[colorinlistoftodos,prependcaption,textsize=tiny]{todonotes}
\usepackage{siunitx}
\usepackage[inline,shortlabels]{enumitem}

\AtlasTitle{%
Search for additional heavy neutral Higgs and gauge bosons in the ditau final state produced in \SI{36}{fb$^{-1}$} of $pp$ collisions at $\sqrt{s} = \SI{13}{\TeV}$ with the ATLAS detector}

\author{The ATLAS Collaboration}

\AtlasRefCode{HIGG-2016-12}

\PreprintIdNumber{CERN-EP-2017-199}

\arXivId{1709.07242}

\HepDataRecord{https://www.hepdata.net/record/ins1624690}

\AtlasJournal{JHEP}
\AtlasJournalRef{JHEP 01 (2018) 055}
\AtlasDOI{10.1007/JHEP01(2018)055}

\AtlasAbstract{%

A search for heavy neutral Higgs bosons and $Z^{\prime}$ bosons is performed using a data sample
corresponding to an integrated luminosity of \SI{36.1}{fb$^{-1}$} from
proton--proton collisions at $\sqrt{s} = \SI{13}{\TeV}$ recorded by the ATLAS
detector at the LHC during 2015 and 2016. The heavy resonance is assumed to
decay to $\tau^+\tau^-$ with at least one tau lepton decaying to final states
with hadrons and a neutrino. The search is performed in the mass range of
0.2--\SI{2.25}{\TeV} for Higgs bosons and 0.2--\SI{4.0}{\TeV} for $Z^{\prime}$
bosons.  The data are in good agreement with the background predicted by the
Standard Model.  The results are interpreted in benchmark scenarios.  In the
context of the hMSSM scenario, the data exclude $\tanb > 1.0$ for $\MA =
\SI{0.25}{\TeV}$ and $\tanb> 42$ for $\MA = \SI{1.5}{\TeV}$ at the 95\%
confidence level.  For the Sequential Standard Model, \ZprimeSSM with
$m_{\Zprime} < \SI{2.42}{\TeV}$ is excluded at 95\% confidence level, while \ZprimeSFM with
$m_{\Zprime} < \SI{2.25}{\TeV}$ is excluded for the non-universal
$G(221)$ model that exhibits enhanced couplings to third-generation fermions.

}

\hypersetup{pdftitle={ATLAS document},pdfauthor={The ATLAS Collaboration}}

\begin{document}

\maketitle

\tableofcontents


\section{Introduction}
\label{sec:intro}

The discovery of a scalar particle~\cite{ATLASHiggsJuly2012, CMSHiggsJuly2012}
at the Large Hadron Collider (LHC)~\cite{LHC} has provided important insight
into the mechanism of electroweak symmetry breaking.  Experimental studies of
the new particle
\cite{Aad:2015mxa,Aad:2015gba,Khachatryan:2014jba,Khachatryan:2014kca,HIGG-2015-07}
demonstrate consistency with the Standard Model (SM) Higgs boson
\cite{ENGLERT,HIGGS,HIGGS2,HIGGS3, Guralnik:1964eu,Kibble:1967sv}. However, it
remains possible that the discovered particle is part of an extended scalar
sector, a scenario that is predicted by a number of theoretical arguments
\cite{Djouadi:2005gj,Branco:2011iw}.

The Minimal Supersymmetric Standard Model
(MSSM)~\cite{Djouadi:2005gj,Fayet:1976et,Fayet:1977yc} is the simplest
extension of the SM that includes supersymmetry. The MSSM requires two Higgs
doublets of opposite hypercharge.  Assuming that CP symmetry is conserved, this
results in one CP-odd ($A$) and two CP-even ($h$, $H$)  neutral Higgs bosons
and two charged Higgs bosons ($H^{\pm}$).  At tree level, the properties of the
Higgs sector in the MSSM depend on only two non-SM parameters, which can be
chosen to be the mass of the CP-odd Higgs boson, $m_A$, and the ratio of the
vacuum expectation values of the two Higgs doublets, $\tan\beta$. Beyond tree
level, a number of additional parameters affect the Higgs sector, the choice of
which defines various MSSM benchmark scenarios. In the 
$m_{h}^{\text{mod}+}$ scenario~\cite{MSSMBenchmarks}, the top-squark mixing parameter is
chosen such that the mass of the lightest CP-even Higgs boson, $m_h$, is close
to the measured mass of the Higgs boson that was discovered at the LHC.  A
different approach is employed in the hMSSM
scenario~\cite{Djouadi:2013uqa,Bagnaschi:2015hka} in which the measured value
of $m_h$ can be used, with certain assumptions, to predict the remaining masses
and couplings of the MSSM Higgs bosons without explicit reference to the soft
supersymmetry-breaking parameters.  The couplings of the MSSM heavy Higgs
bosons to down-type fermions are enhanced with respect to the SM Higgs boson
for large $\tan\beta$ values, resulting in increased branching fractions to
$\tau$-leptons and $b$-quarks, as well as a higher cross section for Higgs
boson production in association with $b$-quarks.  This has motivated a variety
of searches for a scalar boson (generically called $\phi$) in $\tau\tau$ and
$bb$ final states\footnote{The notation $\tau\tau$ and $bb$ is used as
shorthand for $\tau^+\tau^-$ and $b\bar{b}$ throughout this paper.} at
LEP~\cite{LEPLimits}, the
Tevatron~\cite{TevatronLimits1,TevatronLimits2,TevatronLimits3} and the
LHC~\cite{Khachatryan:2014wca, CMS-HIG-14-017,
LHCbHtautau, HIGG-2015-08}.

Heavy \Zprime{} gauge bosons appear in many extensions of the
SM~\cite{Hewett:1989,Cvetic:1995zs,Leike:1998wr,Diener:2010sy,Langacker:2008}
and while they are typically considered to obey lepton universality, this is
not necessarily a requirement. In particular, models in which the \Zprime{}
boson couples preferentially to third-generation fermions may be linked to the
high mass of the top quark~\cite{ZprimeToThirdGen,sfm,G221,TopFlavor} or to
recent indications of lepton flavour universality violation in semi-tauonic $B$
meson decays~\cite{Faroughy:2016osc}.  One such model is the non-universal
$G(221)$ model~\cite{sfm,TopFlavor,G221}, which contains a \ZprimeSFM{} boson
that can exhibit enhanced couplings to tau leptons.  In this model the SM
SU(2) gauge group is split into two parts: one coupling to fermions of the
first two generations and one coupling to third generation fermions. The mixing
between these groups is described by the parameter $\sin^{2}\phi$, with
$\sin^{2}\phi<0.5$ corresponding to enhanced third generation couplings. A
frequently used benchmark model is the Sequential Standard Model (SSM), which
contains a \ZprimeSSM{} boson with couplings identical to the SM $Z$ boson. By
evaluating the impact on the signal sensitivity from changing the \ZprimeSSM{}
couplings, limits on \ZprimeSSM{} can be reinterpreted for a broad range of
models. Indirect limits on \Zprime{} bosons with non-universal flavour
couplings have been derived from measurements at
LEP~\cite{ZprimeMassLimitsFromLEP}.  The most sensitive direct searches for
high-mass  resonances decaying to \ditau final states have been performed by
the ATLAS and CMS collaborations using data collected at $\sqrt{s}=8$ and
$\SI{13}{\TeV}$~\cite{EXOT-2014-05, HIGG-2015-08, CMS-EXO-16-008}.

\begin{figure}[tp]
  \centering
  \subfigure[\label{fig:feyn:ggH}]{\includegraphics[width=0.33\textwidth]{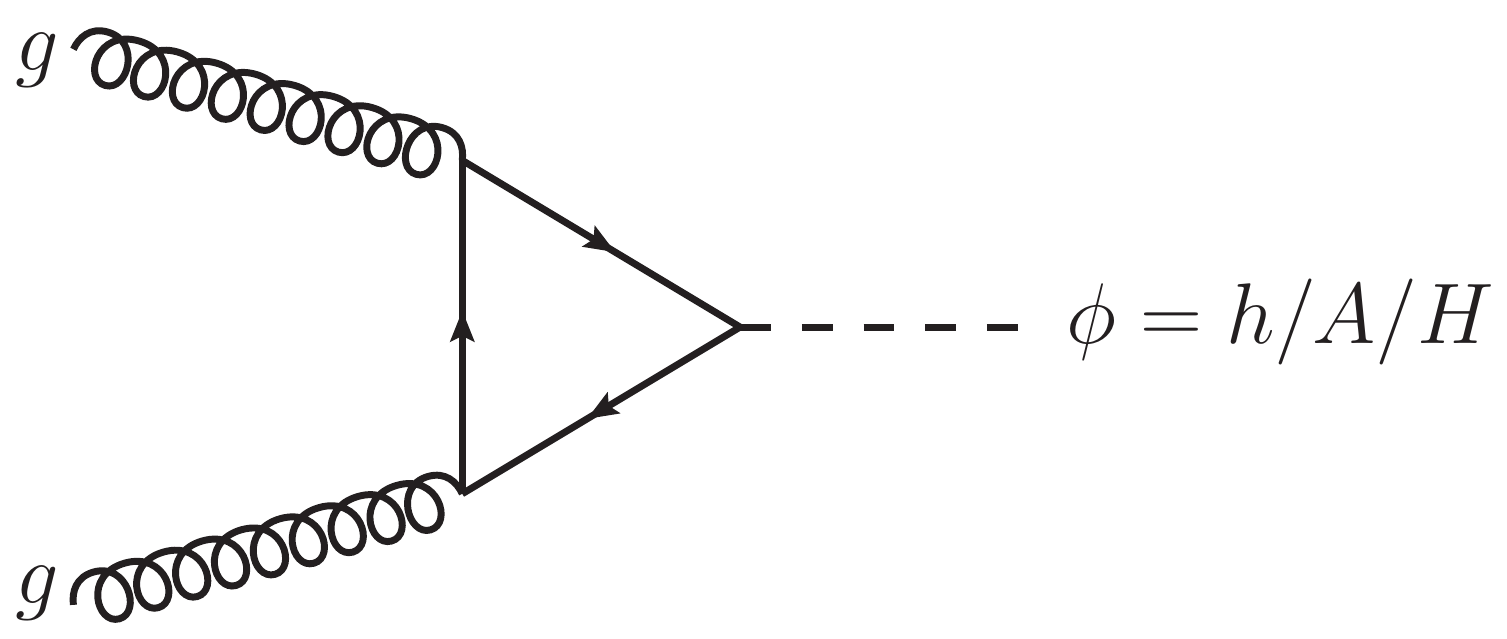} }
  \hspace{1.5cm}  
  \subfigure[\label{fig:feyn:bbH4}]{\includegraphics[width=0.31\textwidth]{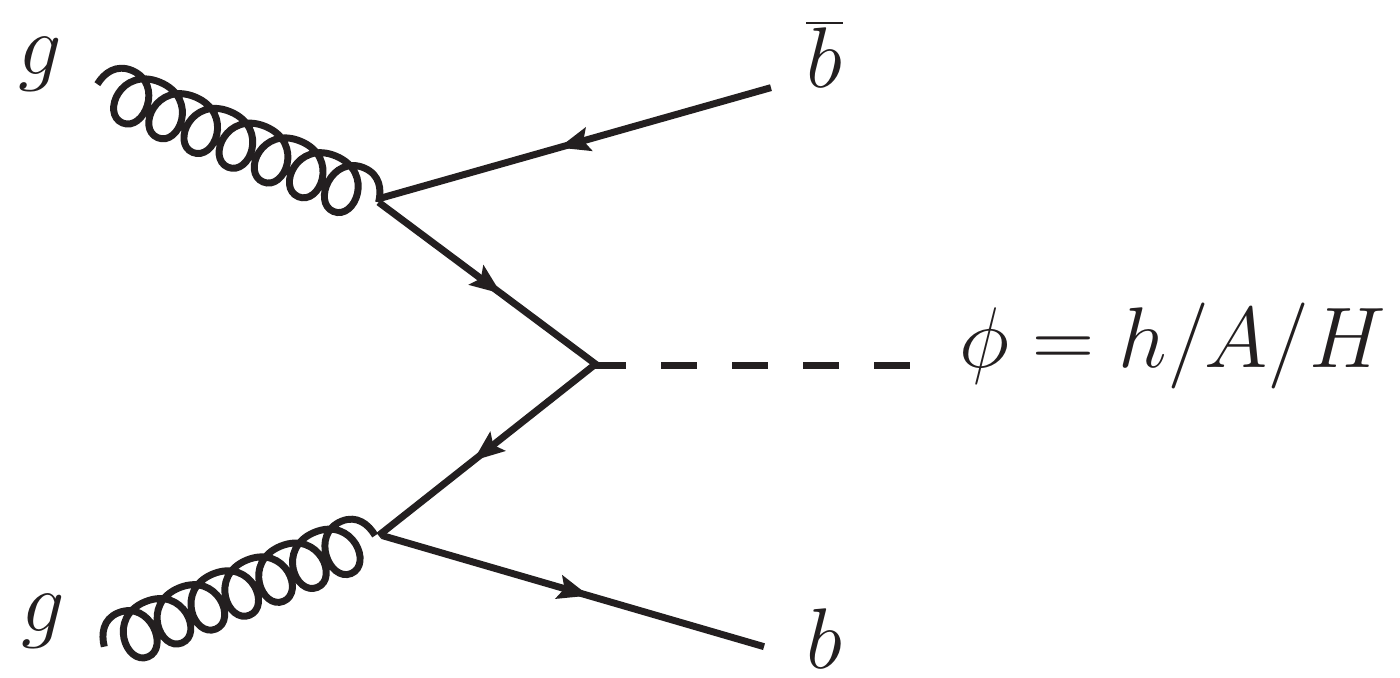} } \\
  \subfigure[\label{fig:feyn:bbH5}]{\includegraphics[width=0.31\textwidth]{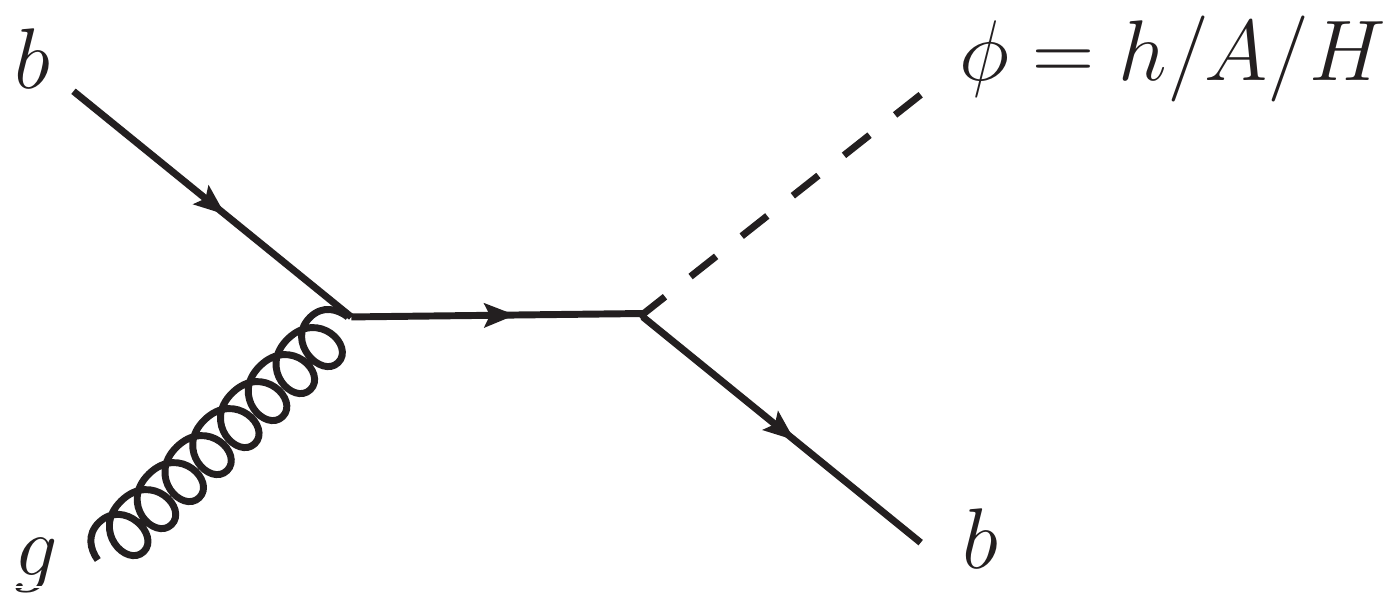} }
  \hspace{2.0cm} 
  \subfigure[\label{fig:feyn:zprime}]{\includegraphics[width=0.26\textwidth]{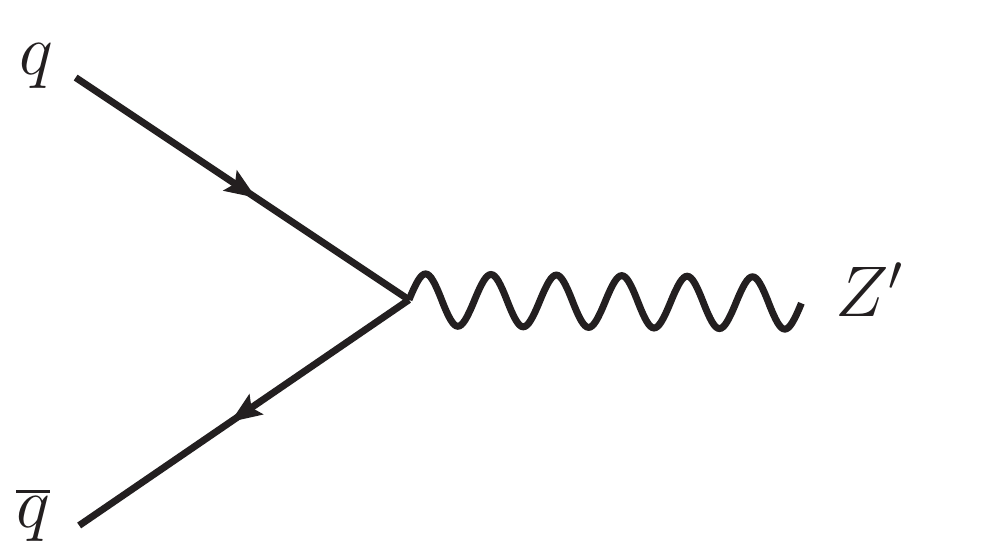} }
  \caption{Lowest-order Feynman diagrams for (a) gluon--gluon fusion and $b$-associated 
production of a neutral MSSM Higgs boson in the (b) four-flavour and (c) five-flavour schemes
and (d) Drell--Yan production of a \Zprime{} boson.}
\label{fig:feynman}
\end{figure}

This paper presents the results of a search for neutral MSSM Higgs bosons as
well as high-mass \Zprime{} resonances in the \ditau decay mode using \lumi of
proton--proton collision data at a centre-of-mass energy of \SI{13}{\TeV}
collected with the ATLAS detector \cite{ATLASDetector} in 2015 and 2016.  The
search is performed in the $\taulep\tauhad$ and $\tauhad\tauhad$ decay modes,
where $\taulep$ represents the decay of a $\tau$-lepton to an electron or a
muon and neutrinos, whereas $\tauhad$ represents the decay to one or more
hadrons and a neutrino.  The search considers narrow resonances\footnote{A
resonance is considered ``narrow'' if the lineshape has no impact on
experimental observables.} with masses of \hmin--\SI{\hmax}{\TeV} and
$\tan\beta$ of 1--58 for the MSSM Higgs bosons. For the \Zprime{} boson search,
a mass range of \zmin--\SI{\zmax}{\TeV} is considered.  Higgs boson production
through gluon--gluon fusion and in association with $b$-quarks is considered
(Figures~\ref{fig:feyn:ggH}--\ref{fig:feyn:bbH5}), with the latter mode
dominating for high $\tan\beta$ values.  Hence, both the $\taulep\tauhad$ and
$\tauhad\tauhad$ channels are split into $b$-tag and $b$-veto categories, based
on the presence or absence of jets tagged as originating from $b$-quarks in the
final state.  Since a \Zprime{} boson is expected to be predominantly produced
via a Drell--Yan process (Figure~\ref{fig:feyn:zprime}), there is little gain
in splitting the data into $b$-tag and $b$-veto categories.  Hence, the
\Zprime{} analysis uses an inclusive selection instead.

The paper is structured as follows. Section~\ref{sec:atlas} provides an
overview of the ATLAS detector.  The event samples used in the analysis,
recorded by the ATLAS detector or simulated using the ATLAS simulation
framework, are reported in Section~\ref{sec:samples}. The event reconstruction
is presented in Section~\ref{sec:reco}. A description of the event selection
criteria is given in Section~\ref{sec:sel}. Section~\ref{sec:bkg} explains the
estimation of background contributions, followed by a description of systematic
uncertainties in Section~\ref{sec:sys}. Results are presented in
Section~\ref{sec:results}, followed by concluding remarks in
Section~\ref{sec:conclusion}.

\newcommand{\AtlasCoordFootnote}{%
ATLAS uses a right-handed coordinate system with its origin at the nominal interaction point
in the centre of the detector and the $z$-axis along the beam pipe.
The $x$-axis points from the interaction point to the centre of the LHC ring,
and the $y$-axis points upwards.
Cylindrical coordinates $(r,\phi)$ are used in the transverse plane,
$\phi$ being the azimuthal angle around the $z$-axis.
The pseudorapidity is defined in terms of the polar angle $\theta$ as $\eta = -\ln \tan(\theta/2)$.
Angular distance is measured in units of $\Delta R \equiv \sqrt{(\Delta\eta)^{2} + (\Delta\phi)^{2}}$.}

\section{ATLAS detector}
\label{sec:atlas}

The ATLAS detector~\cite{ATLASDetector} at the LHC covers nearly the entire solid 
angle around the collision point.\footnote{\AtlasCoordFootnote} It consists of an inner tracking detector surrounded 
by a thin superconducting solenoid, electromagnetic and hadronic calorimeters,
and a muon spectrometer incorporating three large superconducting toroid magnets.
The inner-detector system is immersed in a \SI{2}{\tesla} axial magnetic field
and provides charged-particle tracking in the range $|\eta| < 2.5$.

The high-granularity silicon pixel detector covers the vertex region and typically provides 
four measurements per track.  The innermost layer, known as the insertable B-Layer~\cite{ATLAS-TDR-19}, was added in 2014 and provides 
high-resolution hits at small radius to improve the tracking performance. The pixel detector is 
surrounded by the silicon microstrip tracker, which usually provides four two-dimensional measurement 
points per track. These silicon detectors are complemented by the transition radiation tracker,
which enables radially extended track reconstruction up to $|\eta| = 2.0$. The transition radiation 
tracker also provides electron identification information based on the fraction of hits (typically 
30 in total) above a higher energy-deposit threshold corresponding to transition radiation.

The calorimeter system covers the pseudorapidity range $|\eta| < 4.9$.
Within the region $|\eta|< 3.2$, electromagnetic calorimetry is provided by barrel and
endcap high-granularity lead/liquid-argon (LAr) electromagnetic calorimeters,
with an additional thin LAr presampler covering $|\eta| < 1.8$,
to correct for energy loss in material upstream of the calorimeters.
Hadronic calorimetry is provided by the steel/scintillator-tile calorimeter,
segmented into three barrel structures within $|\eta| < 1.7$, and two copper/LAr hadronic endcap 
calorimeters that cover $1.5 < |\eta| < 3.2$. The solid angle coverage is completed with forward 
copper/LAr and tungsten/LAr calorimeter modules, optimised for electromagnetic and hadronic measurements 
respectively, in the region $3.1 < |\eta| < 4.9$.

The muon spectrometer comprises separate trigger and high-precision tracking chambers measuring the 
deflection of muons in a magnetic field generated by superconducting air-core toroids.
The precision chamber system covers the region $|\eta| < 2.7$ with three layers of monitored drift tubes,
complemented by cathode strip chambers in the forward region, where the background is highest.
The muon trigger system covers the range $|\eta| < 2.4$ with resistive plate chambers in the barrel, and 
thin gap chambers in the endcap regions.

A two-level trigger system is used to select interesting events~\cite{TRIG-2016-01,ATL-DAQ-PUB-2017-001}.
The level-one trigger is implemented in hardware and uses a subset of detector information
to reduce the event rate to a design value of at most \SI{100}{\kHz}.
This is followed by the software-based high-level trigger, which reduces the event rate to \SI{1}{\kHz}.

\section{Data and simulated event samples}
\label{sec:samples}

The results in this paper use proton--proton collision data at a centre-of-mass
energy of $\sqrt{s}=\SI{13}{\TeV}$ collected by the ATLAS detector at the LHC
during 2015 and 2016. The data correspond to a total integrated luminosity of
\lumi after requiring that all relevant components of the ATLAS detector are in good
working condition. Selected events must satisfy criteria designed to reduce
backgrounds from cosmic rays, beam-induced events and calorimeter
noise~\cite{ATLAS-CONF-2015-029}.  They must also contain at least one primary
vertex with at least two associated tracks. The primary vertex is chosen as the
proton--proton vertex candidate with the highest sum of the squared transverse
momenta of the associated tracks.

Simulated events are used to estimate the signal efficiencies and some of the
background contributions. The simulated event samples are normalised using
their theoretical cross sections and the integrated luminosity.  Simulated
events with a heavy neutral MSSM Higgs boson produced via gluon--gluon fusion
and in association with $b$-quarks were generated at next-to-leading order
(NLO) with \POWHEGBOXV{ v2}~\cite{Nason:2004rx,Frixione:2007vw,Powheg3} and
MG5\_aMC@NLO~2.1.2~\cite{Alwall:2014hca,Wiesemann:2014ioa} (using the
four-flavour scheme), respectively.  The CT10~\cite{Lai:2010vv} set of parton
distribution functions (PDFs) was used in the generation of gluon--gluon fusion
events while CT10nlo\_nf4~\cite{Dulat:2015mca} was used to produce the
$b$-associated signal samples.  \PYTHIAV{8.210}~\cite{PYTHIA82} with the
AZNLO~\cite{AZNLO:2014} (A14~\cite{ATL-PHYS-PUB-2014-021}) set of tuned
parameters (tune) was used together with the CTEQ6L1~\cite{Pumplin:2002vw}
(\textsc{NNPDF2.3LO}~\cite{Ball:2012cx}) PDF set for the parton shower
calculation at leading order (LO), underlying event and hadronisation in the
gluon--gluon fusion ($b$-associated) production.  The gluon--gluon fusion
sample was generated assuming SM couplings and underestimates the loop
contribution from $b$-quarks at high \tanb, which can impact the Higgs boson
\pt spectrum.  Generator-level studies indicate this has a negligible impact on
the final mass distribution and only a few percent impact on the signal
acceptance, except for mass hypotheses below \SI{400}{\GeV} where the impact
can be up to 10\%, so the effect is neglected.

The production cross sections and branching fractions for the various MSSM
scenarios are taken from Ref.~\cite{LHCHXSWG2016}.  The cross sections for
gluon--gluon fusion production are calculated using
\SUSHI~\cite{Harlander:2012pb}, including NLO supersymmetric-QCD
corrections~\cite{Spira:1995rr, Harlander:2004tp, Harlander:2005rq,
Degrassi:2010eu, Degrassi:2011vq, Degrassi:2012vt},
next-to-next-to-leading-order (NNLO) QCD corrections for the top
quark~\cite{Harlander:2002wh, Anastasiou:2002yz, Ravindran:2003um,
Harlander:2002vv, Anastasiou:2002wq}, as well as light-quark electroweak
effects~\cite{Aglietti:2004nj, Bonciani:2010ms}.  The $b$-associated production
cross sections in the five-flavour scheme are also calculated using \SUSHI
based on bbh@nnlo~\cite{Harlander:2003ai}, and those for $b$-associated
production in the four-flavour scheme (where $b$-quarks are not considered as
partons) are calculated according to Refs.~\cite{Dittmaier:2003ej,
Dawson:2003kb}.  The final $b$-associated production cross section is obtained
by using the method described in Ref.~\cite{SantanderMatching} to match the
four-flavour and five-flavour scheme cross sections.  The masses and mixing
(and effective Yukawa couplings) of the Higgs bosons are computed with
\FEYNHIGGS~\cite{Heinemeyer:1998yj, Heinemeyer:1998np, Degrassi:2002fi,
Frank:2006yh, Hahn:2013ria, Williams:2011bu} for all scenarios, with the
exception of the hMSSM. In the case of the hMSSM scenario, Higgs masses and
branching fractions are computed using \HDECAY \cite{Djouadi:1997yw,
Djouadi:2006bz}.  Branching fractions for all other scenarios use a combination
of results calculated by \HDECAY, \FEYNHIGGS and
PROPHECY4f~\cite{Bredenstein:2006rh, Bredenstein:2006ha}.

\begin{sloppypar}
The \Zprime{} signal events are modelled with a LO \DY{} sample that is
reweighted with the \TAUSPINNER algorithm~\cite{Czyczula:2012ny,
Kaczmarska:2014eoa, TauSpinnerBSM}, which correctly accounts for spin effects
in the $\tau$-lepton decays.  The \DY{} sample, enriched in events with high
invariant mass, was generated with \PYTHIAV{8.165}~\cite{Pythia8, Pythia} using
the \textsc{NNPDF2.3LO} PDF set and the A14 tune for the parton-shower and
underlying-event parameters.  Interference between the \Zprime{} and the SM
\DY{} production is not included, as it is highly model dependent.
Higher-order QCD corrections are applied to the simulated event samples.  These
corrections to the event yields are made with a mass-dependent rescaling to
NNLO in the QCD coupling constant, as calculated with VRAP
0.9~\cite{Anastasiou:2003ds} and the CT14NNLO PDF set.  Electroweak corrections
are not applied to the \Zprime{} signal samples due to the large model
dependence.
\end{sloppypar}

The \multijet background in both channels is estimated using data, while
non-\multijet backgrounds in which a quark- or gluon-initiated jet is
misidentified as a hadronic tau decay (predominantly \Wjets and \ttbar) are
modelled using data in the \lephad channel and simulation with data-driven
corrections in the \hadhad channel, as described in Section~\ref{sec:bkg}. The
remaining background contributions arise from  \DYjets, \Wjets, \ttbar,
\singletop and \diboson ($WW$, $WZ$ and $ZZ$) production. These contributions
are estimated using the simulated event samples described below.

Events containing \DYjets were generated with \POWHEGBOX{ v2}~\cite{POWHEG}
interfaced to the \PYTHIAV{8.186} parton shower model. The CT10 PDF set was used
in the matrix element. The AZNLO tune was used, with PDF set CTEQ6L1, for the
modelling of non-perturbative effects. 
Photon emission from electroweak vertices and charged leptons was performed 
with \PHOTOSppV{3.52}~\cite{Davidson:2010ew}. The same setup was used to simulate
\Wjets events for background subtraction in the control regions of
the \lephad channel. The \DYjets samples were simulated in slices with different
masses of the off-shell boson. The event yields are corrected with a mass-dependent
rescaling at NNLO in the QCD coupling constant, computed with VRAP 0.9 and the
CT14NNLO PDF set.  Mass-dependent electroweak corrections are computed at NLO
with \MCSANC{1.20}~\cite{Bondarenko:2013nu}, and these include photon-induced
contributions ($\gamma\gamma\rightarrow\ell\ell$ via $t$- and $u$-channel
processes) computed with the MRST2004QED PDF set~\cite{Martin:2004dh}.

\begin{sloppypar}
The modelling of the \Wjets process in the case of the $\tauhad\tauhad$ channel
was done with the \SHERPAV{2.2.0}~\cite{Gleisberg:2008ta} event generator.
Matrix elements were calculated for up to two partons at NLO and four partons
at LO using \COMIX~\cite{Gleisberg:2008fv} and
\OPENLOOPS~\cite{Cascioli:2011va} merged with the \SHERPA parton
shower~\cite{Schumann:2007mg} using the ME+PS@NLO
prescription~\cite{Hoeche:2012yf}. The CT10nlo PDF set was used in conjunction
with dedicated parton shower tuning developed by the \SHERPA authors.  The
\Wjets production is normalised to the NNLO cross sections with
\FEWZ~\cite{Anastasiou:2003ds, Melnikov:2006kv, Gavin:2010az}.
\end{sloppypar}

For the generation of \ttbar or a single top quark in the $Wt$-channel and $s$-channel, the
\POWHEGBOX{ v2} event generator was used with the CT10 PDF set in the matrix element
calculation. Electroweak $t$-channel single-top-quark events were
generated with the \POWHEGBOX{ v1} event generator. This event generator uses the
four-flavour scheme for the NLO matrix elements calculations together with the
fixed four-flavour PDF set CT10f4. For all top processes, top-quark spin
correlations were preserved (for $t$-channel, top quarks were decayed with 
\MADSPIN~\cite{Artoisenet:2012st}). The parton shower, hadronisation, and the
underlying event were simulated using \PYTHIAV{6.428} with the CTEQ6L1 PDF sets
and the corresponding \Perugia 2012 tune~\cite{Skands:2010ak}. The top mass was
set to \SI{172.5}{\GeV}. The \ttbar production sample is normalised to the predicted
production cross section as calculated with the \TOPpp program to NNLO in
perturbative QCD, including soft-gluon resummation to
next-to-next-to-leading-log (NNLL) order (Ref.~\cite{Czakon:2011xx} and references
therein). For the single-top-quark event samples, an approximate
calculation at NLO in QCD for the $s$-channel and 
$t$-channel~\cite{Kant:2014oha,Aliev:2010zk} and an NLO+NNLL calculation for 
the $Wt$-channel~\cite{Kidonakis:2010ux} are used for the normalisation. 

\Diboson processes were modelled using the \SHERPAV{2.1.1} event generator and they
were calculated for up to one ($ZZ$) or no ($WW$, $WZ$) additional partons at NLO
and up to three additional partons at LO using \COMIX and \OPENLOOPS 
merged with the \SHERPA parton shower using the ME+PS@NLO
prescription. The CT10 PDF set was used in conjunction with dedicated parton
shower tuning developed by the \SHERPA authors. The event generator cross sections are
used in this case (already at NLO). In addition, the \SHERPA diboson sample
cross section was scaled down to account for its use of
$\alpha_\textrm{QED} =$ 1/129 rather than 1/132 corresponding to the use of
current PDG parameters as input to the $\mathrm{G}_\mu$ scheme.

Properties of the bottom and charm hadron decays were set with the 
\EVTGENV{v1.2.0} program~\cite{EvtGen} in samples that were not produced with \SHERPA.
Simulated minimum-bias events were overlaid on all simulated samples to include
the effect of multiple proton--proton interactions in the same and neighbouring
bunch crossings (``pile-up''). These minimum-bias events were generated with
\PYTHIAV{8.186}, using the  A2 tune~\cite{ATL-PHYS-PUB-2012-003} and the MSTW2008LO
PDF~\cite{Martin:2009iq}.
Each sample was simulated using the full \GEANTV{4}~\cite{Geant4,ATLASSIM}
simulation of the ATLAS detector, with the exception of the $b$-associated MSSM
Higgs boson signal, for which the \AFII~\cite{ATL-PHYS-PUB-2010-013} fast
simulation framework was used.  Finally, the simulated events are processed
through the same reconstruction software as the data.

\section{Event reconstruction}
\label{sec:reco}

Electron candidates are reconstructed from energy deposits in the
electromagnetic calorimeter associated with a charged-particle track measured
in the inner detector~\cite{PERF-2013-03,PERF-2013-05,ATLAS-CONF-2016-024}.
The electron candidates are required to pass a ``loose'' likelihood-based
identification selection, to have a transverse momentum \mbox{$\pt >
\SI{15}{\GeV}$} and to be in the fiducial volume of the inner detector,
$|\eta|<2.47$.  The transition region between the barrel and endcap
calorimeters ($1.37<|\eta|<1.52$) is excluded. 

Muon candidates are reconstructed in the range $|\eta|<2.5$ by matching tracks
found in the muon spectrometer to tracks found in the inner
detector~\cite{PERF-2015-10}.  The tracks of the muon candidates are re-fitted
using the complete track information from both detector systems. They are
required to have a transverse momentum $\pt>\SI{7}{\GeV}$ and to pass a
``loose'' muon identification requirement.

The selected lepton (electron or muon) in the \lephad channel must then have
$\pt>\SI{30}{\GeV}$ and pass a ``medium'' identification requirement. This lepton is
considered isolated if it meets $\pt$- and $\eta$-dependent isolation criteria utilising
calorimetric and tracking information. The criteria correspond to an efficiency
of 90\% (99\%) for a transverse momentum of $\pt=25~(60)\,\GeV$. The efficiency 
increases with lepton \pt as the requirements are relaxed to account for the 
decreased background from misidentified jets. 

Jets are reconstructed from topological clusters of energy depositions
\cite{Aad:2016upy} in the calorimeter using the anti-$k_t$
algorithm~\cite{AntiKT, IsolationPileUp2}, with a radius parameter value $R
=0.4$. The average energy contribution from pile-up is subtracted according to
the jet area and the jets are calibrated as described in
Ref~\cite{PERF-2016-04}.  They are required to have $\pt > \SI{20}{\GeV}$ and
$|\eta| < 2.5$. To reduce the effect of pile-up, a jet vertex tagger algorithm
is used for jets with $\pt < \SI{60}{\GeV}$ and $|\eta| < 2.4$.  It employs a
multivariate technique based on jet energy, vertexing and tracking variables to
determine the likelihood that the jet originates from or is heavily
contaminated by pile-up~\cite{PERF-2014-03}.  In order to identify jets
containing $b$-hadrons ($b$-jets), a multivariate algorithm is used, which is
based on the presence of tracks with a large impact parameter with respect to
the primary vertex, the presence of displaced secondary vertices and the
reconstructed flight paths of $b$- and $c$-hadrons associated with the
jet~\cite{Aad:2015ydr, ATL-PHYS-PUB-2016-012}.  The algorithm has an average
efficiency of 70\% for $b$-jets and rejections of approximately 13, 56 and 380
for $c$-jets, hadronic tau decays and jets initiated by light quarks or gluons,
respectively, as determined in simulated \ttbar events. 

Hadronic tau decays are composed of a neutrino and a set of visible decay
products (\tauhadvis), typically one or three charged pions and up to two
neutral pions. The reconstruction of the visible decay products is seeded by
jets~\cite{PERF-2013-06}. The \tauhadvis candidates must have $\pt >
25~(45)\,\GeV$ in the \lephad (\hadhad) channel, $|\eta|<2.5$ excluding
$1.37<|\eta|<1.52$, one or three associated tracks and an electric charge of
$\pm 1$. The leading-\pt \tauhadvis candidate in the \lephad channel and the
two leading-\pt \tauhadvis candidates in the \hadhad channel are then selected
and all remaining candidates are considered as jets. A Boosted Decision Tree
(BDT) identification procedure, based on calorimetric shower shapes and
tracking information is used to reject backgrounds from
jets~\cite{ATL-PHYS-PUB-2015-045-fixed,ATLAS-CONF-2017-029}.  Two $\tauhadvis$
identification criteria are used: ``loose'' and ``medium'', specified in
Section~\ref{sec:sel}.  The criteria correspond to efficiencies of about 60\%
(50\%) and 55\% (40\%) in \DYtt events and rejections of about 30 (30)
and 50 (100) in \multijet events, for one-track (three-track) $\tauhadvis$
candidates, respectively.  An additional dedicated likelihood-based veto is
used to reduce the number of electrons misidentified as $\tauhadvis$ in the
\lephad channel, providing 95\% efficiency and a background rejection between
20 and 200, depending on the pseudorapidity of the $\tauhadvis$ candidate.

Geometrically overlapping objects are removed in the following order: 
\begin{enumerate*}[(a)]
\item jets within $\Delta R = 0.2$ of selected \tauhadvis candidates are excluded, 
\item jets within $\Delta R = 0.4$ of an electron or muon are excluded, 
\item any \tauhadvis candidate within $\Delta R = 0.2$ of an electron or muon is excluded,
\item electrons within $\Delta R$ = 0.2 of a muon are excluded.
\end{enumerate*}

The missing transverse momentum, \metvec, is calculated as the
negative vectorial sum of the $\mathbf{p}_\mathrm{T}$ of all fully
reconstructed and calibrated physics
objects~\cite{ATL-PHYS-PUB-2015-027,ATL-PHYS-PUB-2015-023}. This procedure
includes a ``soft term'', which is calculated using the inner-detector
tracks that originate from the hard-scattering vertex but are not associated with
reconstructed objects.

\section{Event selection}
\label{sec:sel}

\subsection{\hadhad channel}
\label{sec:sel:hadhad}

Events in the \hadhad channel are recorded using single-tau triggers with \pt
thresholds of 80, 125 or \SI{160}{\GeV}, depending on the data-taking period.
Events must contain at least two \tauhadvis candidates with $\pt >
\SI{65}{\GeV}$ and no electrons or muons. The leading-\pt \tauhadvis candidate
must be geometrically matched to the trigger signature and must exceed the
trigger \pt threshold by \SI{5}{\GeV}. The leading and sub-leading \tauhadvis
candidates must satisfy the ``medium'' and ``loose'' identification criteria,
respectively.  They must also have opposite electric charge and be back to back
in the transverse plane: $|\Delta\phi(\ptvec^{\tau_1},\ptvec^{\tau_2})| >
\SI{2.7}{rad}$, as tau leptons from the decay of heavy neutral resonances are
typically produced back to back in the transverse plane.  The signal acceptance
times efficiency for this selection (calculated with respect to all possible
\ditau final states) varies between 1\% and 7\% for signals with masses of
\SI{0.35}{\TeV} or higher. The maximum occurs for signals with masses of around
\SI{0.9}{\TeV}, degradations occur at lower masses due to the \tauhadvis \pt
thresholds and at higher masses due to the \tauhadvis
reconstruction and identification efficiencies.  A summary of the selection is given in
Table~\ref{tab:regions:hadhad} of Section~\ref{sec:bkg}.

\subsection{\lephad channel}
\label{sec:sel:lephad}

\begin{sloppypar}
Events in the \lephad channel are recorded using single-electron and single-muon
triggers with \pt thresholds ranging from 20 to \SI{140}{\GeV} and various 
isolation criteria. The events must contain at least one \tauhadvis candidate
passing the medium identification, exactly one isolated lepton (from here on
referred to as $\ell$) that is geometrically matched to the trigger signature
(implying $|\eta|<2.4$ in the \muhad channel), and no additional
reconstructed leptons.  The identified \tauhadvis candidate must have
$|\eta|<2.3$ to reduce background from misidentified electrons. The isolated
lepton and identified \tauhadvis candidate must have opposite electric charge
and be back to back in the transverse plane: $|\Delta\phi(\ptvec^{\ell},\ptvec^{\tauhadvis})|>
\SI{2.4}{rad}$. To reduce background
from \Wjets production, the transverse mass of the isolated lepton and the
missing transverse momentum,
\begin{equation*} 
\mt(\ptvec^{\ell},\metvec) \equiv \sqrt{2 \pt^{\ell}\met \left [ 1 - \cos\Delta\phi(\ptvec^{\ell},
\metvec) \right ] }\,, 
\end{equation*}
must be less than $\SI{40}{\GeV}$. To reduce background from \Zee production in
the \ehad channel, events where the isolated lepton and identified \tauhadvis
candidate have an invariant mass between 80 and \SI{110}{\GeV} are rejected.
The signal acceptance times efficiency for this selection  
also varies between 1\% and 7\%, but the maximum 
occurs at lower masses due to the lower \pt thresholds on the tau decay products. 
A summary of the selection is given 
in Table~\ref{tab:regions:lephad} of Section~\ref{sec:bkg}. 
\end{sloppypar}

\subsection{Event categories}
\label{sec:sel:categories}

Events satisfying the selection criteria in the \lephad and \hadhad channels are 
categorised to exploit the different production modes in the MSSM. Events containing 
at least one $b$-tagged jet enter the \btag category, while events containing 
no $b$-tagged jets enter the \bveto category. The categorisation is not used 
for the \Zprime search.

\subsection{\Ditau mass reconstruction}
\label{sec:sel:massreco}

The \ditau mass reconstruction is important for achieving good separation
between signal and background. However, \ditau mass reconstruction is challenging due
to the presence of neutrinos from the $\tau$-lepton decays. 
Therefore, the mass reconstruction used for both the $\tauhad\tauhad$ and $\taulep\tauhad$ channels
is the total transverse mass, defined as:
\begin{equation*}
\mttot \equiv \sqrt{ (p_\mathrm{T}^{\tau_1} + p_\mathrm{T}^{\tau_2} + \MET)^2 
 - (\mathbf{p}_\mathrm{T}^{\tau_1} + \mathbf{p}_\mathrm{T}^{\tau_2} + \mathbf{\MET})^2 }\,,
\end{equation*}
where $\mathbf{p}_\mathrm{T}^{\tau_1}$ and $\mathbf{p}_\mathrm{T}^{\tau_2}$ are the momenta 
of the visible tau decay products (including \tauhad and \taulep) projected into the transverse plane.
More complex mass reconstruction techniques were investigated, but they did not improve the expected sensitivity.

\section{Background estimation}
\label{sec:bkg}

The dominant background contribution in the \hadhad channel is from \multijet
production, which is estimated using a data-driven technique, described in
Section~\ref{sec:bkg:hadhad}.  Other important background contributions come
from \DYtt production at high \mttot in the \bveto category, \ttbar production
in the \btag category, and to a lesser extent \Wlnujets, \singletop, \diboson
and \DYlljets production.  These contributions are estimated using simulation.
Corrections are applied to the simulation to account for mismodelling of the
trigger, reconstruction, identification and isolation efficiencies, the
electron to \tauhadvis misidentification rate and the momentum scales and
resolutions. To further improve the modelling in the \hadhad channel, events in
the simulation that contain quark- or gluon-initiated jets (henceforth called
jets) that are misidentified as \tauhadvis candidates are weighted by
fake-rates measured in \Wjets and \ttbar control regions in data.

The dominant background contribution in the \lephad channel arises from
processes where the \tauhadvis candidate originates from a jet. This
contribution is estimated using a data-driven technique similar to the 
\hadhad channel, described
in Section~\ref{sec:bkg:lephad}. The events are divided into those where the
selected lepton is correctly identified, predominantly from \Wjets (\ttbar)
production in the \bveto (\btag) channel, and those where the selected lepton
arises from a jet, predominantly from \multijet production.  Backgrounds where
both the \tauhadvis and lepton candidates originate from electrons, muons or
taus (\tworeal) arise from \DYtt production in the \bveto category and \ttbar
production in the \btag category, with minor contributions from \DYll, \diboson
and \singletop production. These contributions are estimated using simulation.
To help constrain the normalisation of the \ttbar contribution, a
control region rich in \ttbar events (CR-T) is defined and included in the
statistical fitting procedure. The other major background contributions can be
adequately constrained in the signal regions. Events in this control region
must pass the signal selection for the \btag category, but the
$\mT(\ptvec^{\ell},\metvec)$ selection is replaced by
$\mT(\ptvec^{\ell},\metvec) > 110\, (100)\,\GeV$ in the \ehad (\muhad) channel.
The tighter selection in the \ehad channel is used to help suppress the larger
\multijet contamination. The region has $\sim$90\% \ttbar purity.

\subsection{Jet background estimate in the \hadhad channel}
\label{sec:bkg:hadhad}
The data-driven technique used to estimate the dominant \multijet background in
the \hadhad channel is described in Section~\ref{sec:hadhad:multijet}.  The
method used to weight simulated events to estimate the remaining background
containing events with \tauhadvis candidates that originate from jets is
described in Section~\ref{sec:hadhad:non-multijet}. 
A summary of the signal, control and fakes regions used in these methods 
is provided in Table~\ref{tab:regions:hadhad}. The associated
uncertainties are discussed in Section~\ref{sec:sys:data}.

\begin{table}[tp]
\small
\begin{center}

\caption{Definition of signal, control and fakes regions used in the \hadhad channel. The symbol 
$\tau_1$ ($\tau_2$) represents the leading (sub-leading) \tauhadvis candidate. } 
\label{tab:regions:hadhad}

\begin{tabular}{ll}
\toprule
Region & Selection \\
\midrule
SR    & $\tau_{1}$ (trigger, medium), $\tau_{2}$ (loose), $q(\tau_1)\times q(\tau_2)<0$, $|\Delta\phi(\ptvec^{\tau_1},\ptvec^{\tau_2})|>2.7$ \\
\midrule
CR-1  & Pass SR except: $\tau_2$ (fail loose) \\
DJ-FR & jet trigger, $\tau_1$+$\tau_2$ (no identification), $q(\tau_1)\times q(\tau_2)<0$, $|\Delta\phi(\ptvec^{\tau_1},\ptvec^{\tau_2})|>2.7$, $\pt^{\tau_2}/\pt^{\tau_1} > 0.3$ \\
W-FR & $\mu$ (trigger, isolated), $\tau_1$ (no identification), $|\Delta\phi(\ptvec^{\mu},\ptvec^{\tau_1})|>2.4$, $\mT(\ptvec^{\mu},\metvec)>\SI{40}{\GeV}$ \\
& \bveto category only \\
T-FR & Pass W-FR except: \btag category only \\
\bottomrule
\end{tabular}
\end{center}
\end{table}

\subsubsection{\Multijet events}
\label{sec:hadhad:multijet}

The contribution of \multijet events in the signal region (\sr) of the \hadhad
channel is estimated using events in two control regions (\failcr and \djcr).
Events in \failcr must pass the same selection as \sr, but the sub-leading
\tauhadvis candidate must fail \tauhadvis identification. The non-\multijet
contamination in this region, $N_\mathrm{non-MJ}^\mathrm{\failcr}$, amounts to $\sim$$1.6\%$
($\sim$$7.0\%$) in the \bveto (\btag) channel, and is subtracted using
simulation. Events in \djcr (the \dijet fakes-region) are used to measure
fake-factors (\fDJ), which are defined as the ratio of the number
of \tauhadvis that pass to those that fail the identification. The 
fake-factors are used to weight the events in \failcr to estimate
the \multijet contribution: 
\begin{equation*}
N_\mathrm{\multijet}^\mathrm{SR} = \fDJ \times 
\left ( N_\mathrm{data}^\mathrm{\failcr} 
      - N_\mathrm{non-MJ}^\mathrm{\failcr}  \right )\,.
\end{equation*}

\begin{figure}[tp]
\begin{center}
\includegraphics[width=0.49\columnwidth]{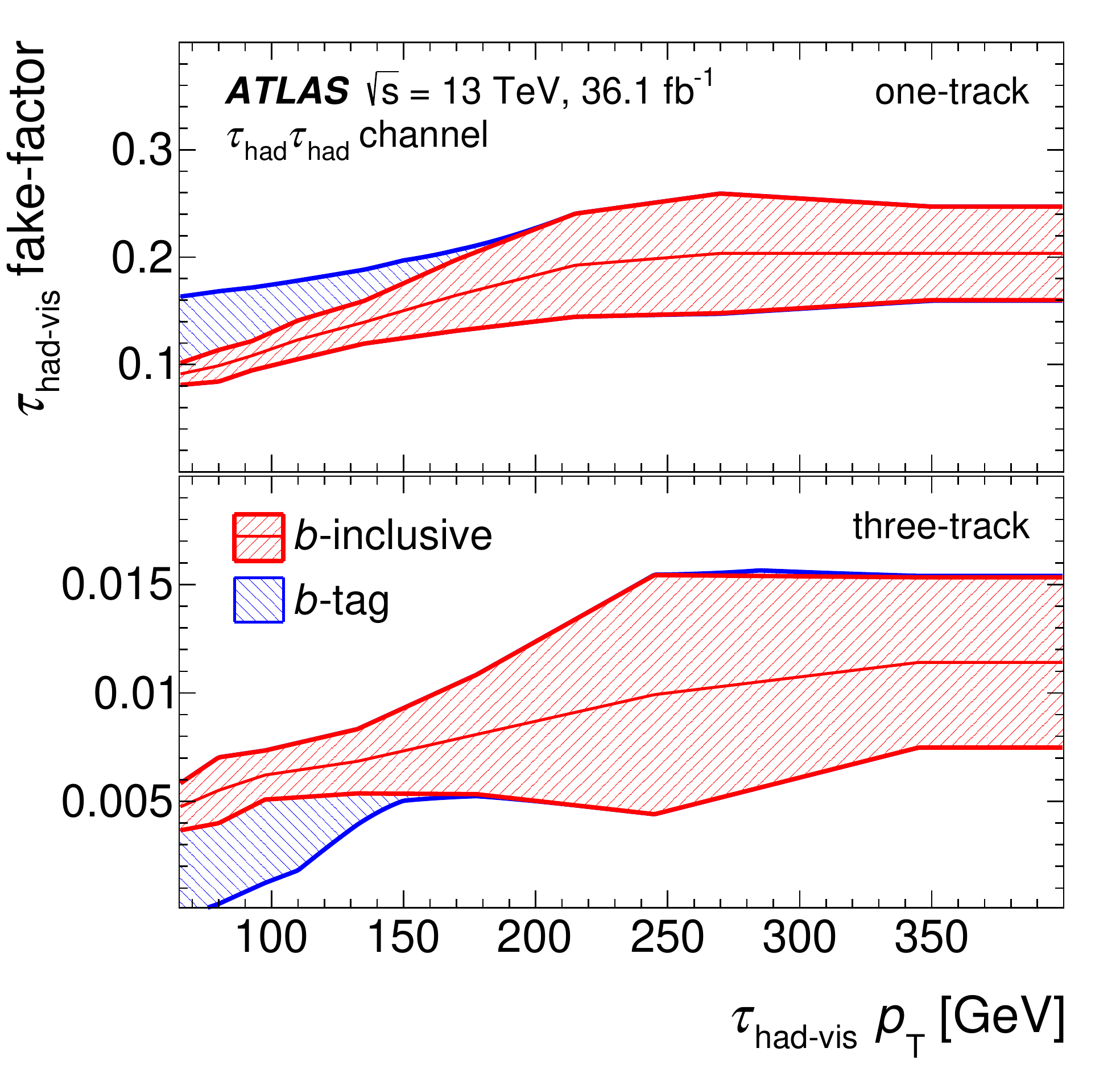}
\caption{The \tauhadvis identification fake-factors in the \hadhad channel.
The red band indicates the total uncertainty when used with a 
$b$-inclusive or $b$-veto selection. The blue band indicates the 
additional uncertainty when used with a \btag selection.}
\label{fig:hadhad_fakefactors}
\end{center}
\end{figure}

The selection for the \djcr is designed to be as similar to the signal
selection as possible, while avoiding contamination from \tauhadvis. Events are
selected by single-jet triggers with \pt thresholds ranging from
\SIrange{60}{380}{\GeV}, with all but the highest-threshold trigger being
prescaled.  They must contain at least two \tauhadvis candidates, where the
leading candidate has $\pt>\SI{85}{\GeV}$ and also exceeds the trigger
threshold by $10\%$, and the sub-leading candidate has $\pt>\SI{65}{\GeV}$. The
\tauhadvis candidates must have opposite charge sign, be back to back in the
transverse plane, $|\Delta\phi(\ptvec^{\tau_1},\ptvec^{\tau_2})| > 2.7$ rad and
the \pt of the sub-leading \tauhadvis must be at least 30\% of the leading
\tauhadvis \pt.  The fake-factors are measured using the sub-leading \tauhadvis
candidate to avoid trigger bias and to be consistent with their application in
\failcr.  They are parameterised as functions of the sub-leading \tauhadvis \pt
and the sub-leading \tauhadvis track multiplicity.  The purity of \multijet
events that pass the \tauhadvis identification is 98--99\% (93--98\%) for the
\bveto (\btag) categories.  The non-\multijet contamination is subtracted using
simulation.  The fake-factors are shown in Figure~\ref{fig:hadhad_fakefactors}.

\subsubsection{Non-\multijet events}
\label{sec:hadhad:non-multijet}

In the \hadhad{} channel, backgrounds originating from jets that are
misidentified as \tauhadvis in processes other than \multijet{} production
(predominantly \Wjets{} in the \bveto and \ttbar in the \btag categories) are
estimated using simulation.  Rather than applying the \tauhadvis identification
to the simulated jets, they are weighted by \frate{}s as in
Ref.~\cite{EXOT-2014-05}. This not only ensures the correct \frate{}, but also
enhances the statistical precision of the estimate, as events failing the
\tauhadvis identification are not discarded.  The \frate{} for the sub-leading
\tauhadvis candidate is defined as the ratio of the number of candidates that
pass the identification to the total number of candidates.  The \frate{} for
the leading \tauhadvis candidate is defined as the ratio of the number of
candidates that pass the identification and the single-tau trigger requirement
to the total number of candidates. 

The fake-rates applied to \ttbar and single-top-quark events are calculated
from a fakes region enriched in \ttbar events (T-FR), while the fake-rates for
all other processes are calculated in a fakes region enriched in \Wjets events
(W-FR). Both T-FR and W-FR use events selected by a single-muon trigger with a
\pt threshold of \SI{50}{\GeV}. They must contain exactly one isolated muon
with $\pt>\SI{55}{\GeV}$ that fired the trigger, no electrons and at least one
\tauhadvis candidate with $\pt>\SI{50}{\GeV}$. The events must also satisfy
$|\Delta\phi(\ptvec^{\mu},\ptvec^{\tauhadvis})|>\SI{2.4}{rad}$ and
$\mT(\ptvec^{\mu},\metvec)>\SI{40}{\GeV}$.  The events are then categorised
into \btag and \bveto categories, defining T-FR and W-FR, respectively.
Backgrounds from non-\ttbar (non-\Wjets) processes are subtracted from T-FR
(W-FR) using simulation.  The fake-rates are measured using the leading-\pt
\tauhadvis candidate and are parameterised as functions of the \tauhadvis \pt
and track multiplicity.

\subsection{Jet background estimate in the \lephad channel}
\label{sec:bkg:lephad}

\begin{figure}[tp]
\begin{center}
\includegraphics[width=0.80\columnwidth]{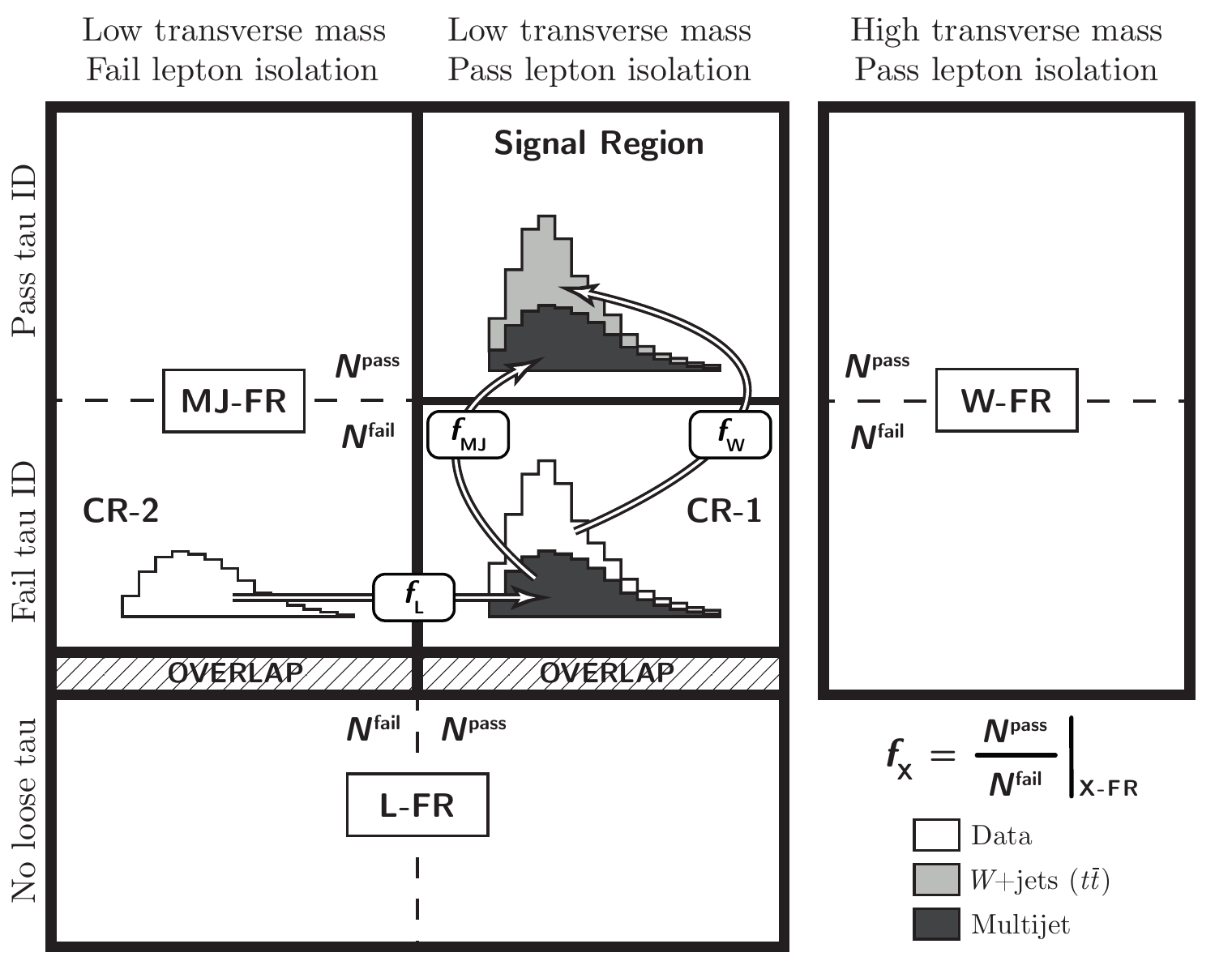}

\caption{
Schematic of the fake-factor background estimation in the \lephad channel. The
fake-factors, $f_\textrm{X}$ (X = MJ, W, L), are defined as the ratio of events
in data that pass/fail the specified selection requirements, measured in the
fakes-regions: MJ-FR, W-FR and L-FR, respectively.  The \multijet contribution
is estimated by weighting events in \mjcrtwo by the product of \flep and \fMJ.
The contribution from \Wjets and \ttbar events where the \tauhadvis candidate
originates from a jet is estimated by subtracting the \multijet contribution
from \failcr and then weighting by \fW.  There is a small overlap of events 
between L-FR and the CR-1 and CR-2 regions. The contribution where both the
selected \tauhadvis and lepton originate from leptons is estimated using
simulation (not shown here).
}
\label{fig:lephad_crs}
\end{center}
\end{figure}

The background contribution from events where the \tauhadvis candidate
originates from a jet in the \lephad channel is estimated using a data-driven
method, which is similar to the one used to estimate the \multijet contribution
in the \hadhad channel. Events in the control region \failcr must pass the same
selection as the \lephad \sr, but the \tauhadvis candidate must fail \tauhadvis
identification. These events are weighted to estimate the jet background in
\sr, but the weighting method must be extended to account for the fact that
\failcr contains both \multijet and \Wjets (or \ttbar) events, which have
significantly different fake-factors.  This is mainly due to a different
fraction of quark-initiated jets, which are typically more narrow and produce
fewer hadrons than gluon-initiated jets, and are thus more likely to pass the
\tauhadvis identification. The procedure, depicted in
Figure~\ref{fig:lephad_crs}, is described in the following. A summary of the
corresponding signal, control and fakes regions is provided in
Table~\ref{tab:regions:lephad}. The associated uncertainties are discussed in
Section~\ref{sec:sys:data}.

\begin{table}[tp]
\small
\begin{center}

\caption{Definition of signal, control and fakes regions used in the \lephad channel. The symbol 
$\ell$ represents the selected electron or muon candidate and $\tau_1$ 
represents the leading \tauhadvis candidate. } 
\label{tab:regions:lephad}

\begin{tabular}{ll}
\toprule
Region & Selection \\
\midrule
SR    & $\ell$ (trigger, isolated), $\tau_1$ (medium), $q(\ell)\times q(\tau_1)<0$, $|\Delta\phi(\ptvec^{\ell},\ptvec^{\tau_1})|>2.4$, \\
& $\mT(\ptvec^{\ell},\metvec)<\SI{40}{\GeV}$, veto $80 < m(\pvec^{\ell},\pvec^{\tau_1}) < \SI{110}{\GeV}$ (\ehad channel only) \\
\midrule
CR-1  & Pass SR except: $\tau_1$ (very-loose, fail medium) \\
CR-2  & Pass SR except: $\tau_1$ (very-loose, fail medium), $\ell$ (fail isolation) \\
MJ-FR & Pass SR except: $\tau_1$ (very-loose), $\ell$ (fail isolation) \\
W-FR  & Pass SR except: 70\,(60)$ < \mT(\ptvec^{\ell},\metvec) < \SI{150}{\GeV}$ in \ehad (\muhad) channel \\
CR-T  & Pass SR except: $\mT(\ptvec^{\ell},\metvec)> 110\,(100)\,\GeV$ in the \ehad (\muhad) channel, \btag category only \\
L-FR  & $\ell$ (trigger, selected), jet (selected), no loose $\tauhadvis$, $\mT(\ptvec^{\ell},\metvec) < \SI{30}{\GeV}$ \\
\bottomrule
\end{tabular}
\end{center}
\end{table}

\subsubsection{\Multijet events}
\label{sec:lephad:multijet}

The \multijet contributions in both \failcr ($N_\mathrm{\multijet}^\mathrm{\failcr}$) 
and \sr ($N_\mathrm{\multijet}^\mathrm{\sr}$) are estimated from events where the 
\tauhadvis fails identification and the selected lepton fails isolation (\mjcrtwo). 
The non-\multijet background is subtracted using simulation and the 
events are weighted first by the lepton-isolation fake-factor (\flep), yielding 
$N_\mathrm{\multijet}^\mathrm{\failcr}$, and then by the \multijet tau fake-factor (\fMJ):
\begin{align*}
N_\mathrm{\twofake}^\mathrm{\failcr}
&= \flep \times 
\left ( N_\mathrm{data}^\mathrm{\mjcrtwo} 
      - N_\mathrm{non-MJ}^\mathrm{\mjcrtwo}  \right )\,,\\
N_\mathrm{\twofake}^\mathrm{\sr} 
&= \fMJ \times N_\mathrm{\twofake}^\mathrm{\failcr}\,.
\end{align*}
The fake-factor \fMJ is measured in the \multijet fakes-region (\mjcr) defined in
Section~\ref{sec:lephad:tidff} and the fake-factor \flep is measured in the 
lepton fakes-region (\flcr) defined in Section~\ref{sec:lephad:lepff}.

\subsubsection{Non-\multijet events}
\label{sec:lephad:non-multijet}

The contribution from \Wjets (and \ttbar) events  where the \tauhadvis candidate 
originates from a jet is estimated from events in
\failcr that remain after subtracting the \multijet contribution and the \tworeal contribution (estimated
using simulation). The events are weighted by the \Wjets tau fake-factor (\fW):
\begin{equation*}
N_\mathrm{\Wjets}^\mathrm{SR} = \fW \times 
\left ( N_\mathrm{data}^\mathrm{\failcr} 
      - N_\mathrm{\twofake}^\mathrm{\failcr}
      - N_\mathrm{\tworeal}^\mathrm{\failcr} \right )\,.
\end{equation*}
The fake-factor \fW is measured in the \Wjets fakes-region (\wcr) defined in Section~\ref{sec:lephad:tidff}.

\subsubsection{Tau identification fake-factors}
\label{sec:lephad:tidff}

\begin{figure}[tp]
\begin{center}
\subfigure[\tauhadvis fake-factors\label{fig:lephad_ffonly}]{\includegraphics[width=0.45\columnwidth]{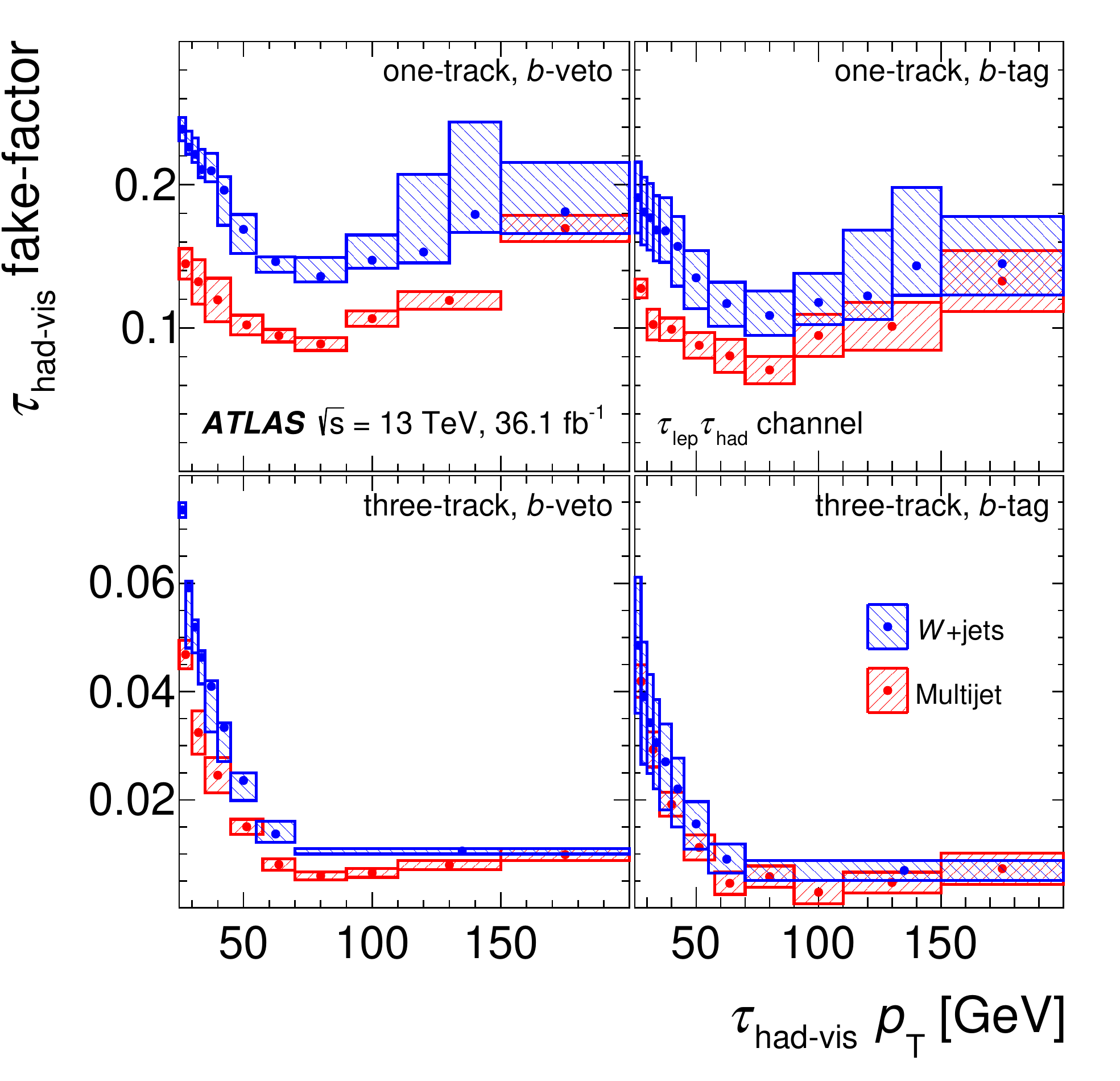}}
\subfigure[$|\Delta\phi(\ptvec^{\tauhadvis}, \metvec)|$ correction\label{fig:lephad_wcorr}]{\includegraphics[width=0.45\columnwidth]{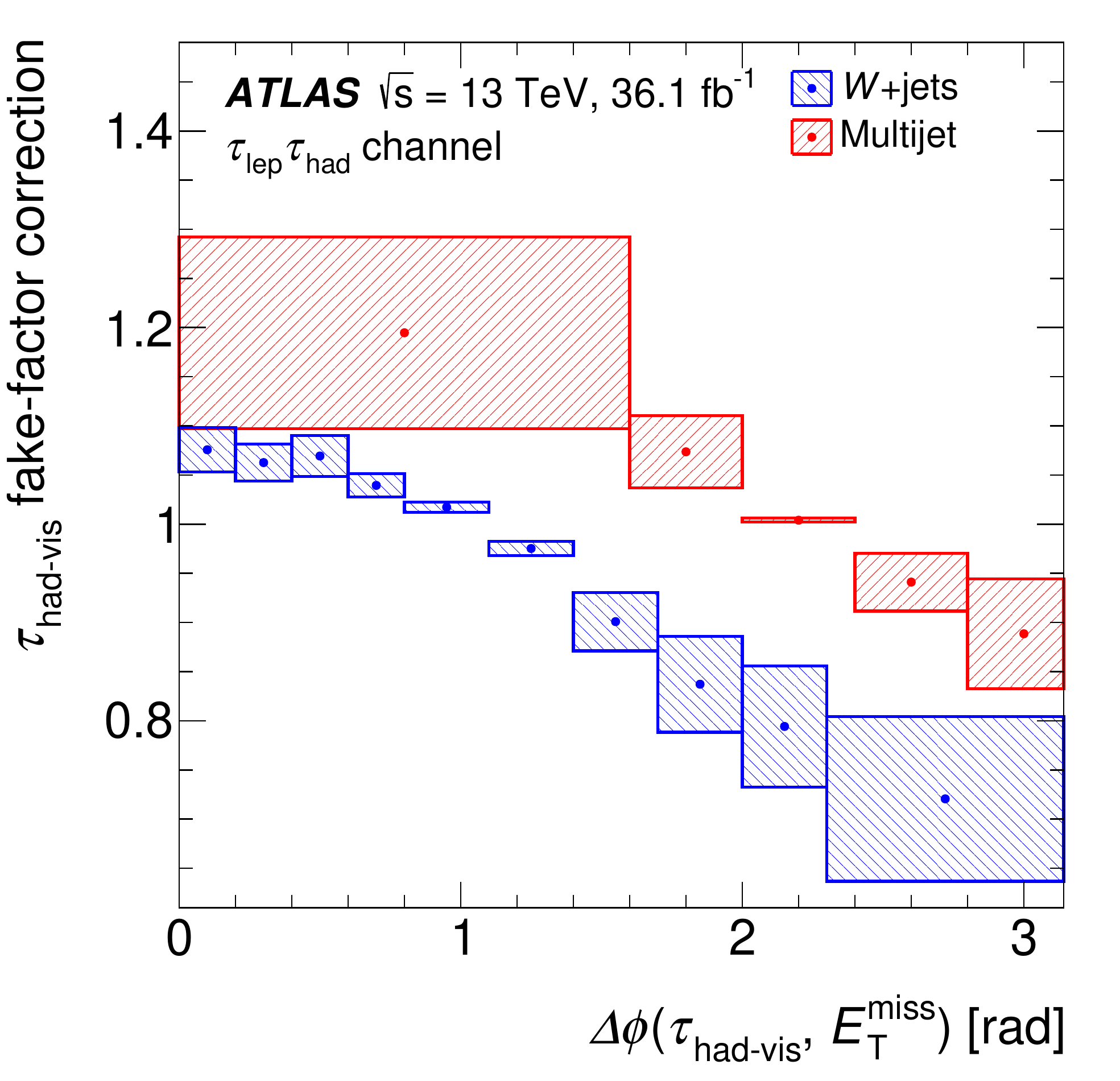}}
\caption{The \tauhadvis identification fake-factors and the sequential $|\Delta\phi(\ptvec^{\tauhadvis},\metvec)|$ 
correction in the \lephad channel. 
The \multijet fake-factors are for the 2016 dataset only.
The bands include all uncertainties. 
}
\label{fig:lephad_fakefactors}
\end{center}
\end{figure}

\begin{sloppypar}
Both \fW and \fMJ are parameterised as functions of \tauhadvis \pt, \tauhadvis
track multiplicity and $|\Delta\phi(\ptvec^{\tauhadvis},\metvec)|$.  The
$|\Delta\phi(\ptvec^{\tauhadvis},\metvec)|$ dependence is included to
encapsulate correlations between the \tauhadvis identification and energy
response, which impact the \metvec calculation. Due to the limited size of the
control regions, the $|\Delta\phi(\ptvec^{\tauhadvis},\metvec)|$ dependence is
extracted as a sequential correction and is only applied in the \bveto channel.
The selection for \wcr and \mjcr are the same as for \sr with modifications
described in the following.  The medium \tauhadvis identification criterion is
replaced by a very loose criterion with an efficiency of about 99\% for
\tauhadvis and a rejection of about 2 (3) for one-track (three-track) jets.
Events passing the medium identification criterion enter the fake-factor
numerators, while those failing enter the denominators.  The very loose
identification reduces differences between \fW and \fMJ, as it tends to reject
gluon-initiated jets, enhancing the fraction of quark-initiated jets in \wcr
and \mjcr. This selection is also applied consistently to \failcr.  A
comparison of the two fake-factors and their respective
$|\Delta\phi(\ptvec^{\tauhadvis},\metvec)|$ corrections are shown in
Figures~\ref{fig:lephad_ffonly} and \ref{fig:lephad_wcorr}.
\end{sloppypar}

In \mjcr, the selected lepton must fail isolation. The \multijet purity for
events that pass the \tauhadvis identification in this region is $\sim$88\% for
the \bveto category and $\sim$93\% for the \btag category.   All non-\multijet
contamination is subtracted from \mjcr using simulation.  The fake-factor \fMJ is further
split by category (\bveto, \btag) and by data-taking period (2015, 2016) to
account for changing isolation criteria in the trigger that affect \mjcr
differently to \sr.

In the \wcr, the $\mT(\ptvec^{\ell},\metvec)$ criterion is replaced by $70 (60)
< \mT(\ptvec^{\ell},\metvec) < \SI{150}{\GeV}$ in the \ehad (\muhad) channel.
The purity of \Wjets events that pass the \tauhadvis identification is
$\sim$85\% in the \bveto category. The \btag category is dominated by \ttbar
events, but the purity of events where the \tauhadvis candidate originates from
a jet is only $\sim$40\% due to the significant fraction of \tauhadvis from $W$
boson decays.  The \multijet and \tworeal backgrounds are subtracted from \wcr
analogously to \failcr in the \Wjets estimate.  Due to the large \tauhadvis
contamination in the \btag region, \fW is not split by category, but the \bveto
parameterisation is used in the \btag region, with a \pt-independent correction
factor of 0.8 (0.66) for one-track (three-track) \tauhadvis. The correction
factor is obtained from a direct measurement of the fake-factors in \btag
events. 

\subsubsection{Lepton isolation fake-factor}
\label{sec:lephad:lepff}

\begin{sloppypar}

The fake-factor \flep is measured in \flcr, which must have exactly one
selected lepton, \mbox{$\mT(\ptvec^{\ell},\metvec)<\SI{30}{\GeV}$} and no
\tauhadvis candidates passing the loose identification but rather at least one
selected jet (not counting the $b$-tagged jet in the \btag region). The
selection is designed to purify \multijet events while suppressing \Wjets and
\ttbar events. Events where the selected lepton passes (fails) isolation enter
the \flep numerator (denominator).  All non-\multijet contributions are
subtracted using simulation. The fake-factors are parameterised as a function
of lepton $|\eta|$, and are further split by lepton type (electron, muon),
category (\bveto, \btag) and into two regions of muon \pt, due to differences
in the isolation criteria of the low- and high-\pt triggers in the \muhad
channel. 
\end{sloppypar}

\section{Systematic uncertainties}
\label{sec:sys}

Uncertainties affecting the simulated signal and background
contributions are discussed in Section~\ref{sec:sys:sim}. These include uncertainties associated with the
determination of the integrated luminosity, the detector simulation,  the
theoretical cross sections and the modelling from the event generators.  
Uncertainties associated with the data-driven background estimates are
discussed in Section~\ref{sec:sys:data}.

\subsection{Uncertainties in simulation estimates}
\label{sec:sys:sim}

The uncertainty in the combined 2015+2016 integrated luminosity is 3.2\%, which
affects all simulated samples. It is derived, following a methodology similar
to that detailed in Ref.~\cite{DAPR-2013-01}, from a preliminary calibration of
the luminosity scale using $x$--$y$ beam-separation scans performed in August
2015 and May 2016.  The uncertainty related to the overlay of pile-up events is
estimated by varying the average number of interactions per bunch crossing by
9\%.  The uncertainties related to the detector simulation manifest themselves through the
efficiency of the reconstruction, identification and triggering algorithms, and
the energy scale and resolution for electrons, muons, \tauhadvis, ($b$-)jets
and the \metvec soft term. These uncertainties are considered for all simulated
samples; their impact is taken into account when estimating signal and 
background contributions and when subtracting contamination from regions 
in the data-driven estimates.  The effects of the particle energy-scale uncertainties
are propagated to \metvec.  The uncertainty in the \tauhadvis identification
efficiency as determined from measurements of \Ztautau events is 5--6\%. 
At high \pt{}, there are no abundant sources of
real hadronic tau decays from which an efficiency measurement could be made.
Rather, the tau identification is studied in high-\pt{} dijet events as a
function of the jet \pt{}, which indicates that there is no degradation in the
modelling of the detector response as a function of the \pt{} of tau
candidates.  Based on the limited precision of these studies, an additional
uncertainty of 20\%/\TeV{} (25\%/\TeV) for one-track (three-track) \tauhadvis candidates with
$\pt>\SI{150}{\GeV}$ is assigned.  The \tauhadvis trigger efficiency uncertainty 
is 3--14\%.
The uncertainty in the \tauhadvis energy scale is 2--3\%. The probability for
electrons to be misidentified as \tauhadvis is measured with a precision of
3--14\%~\cite{ATLAS-CONF-2017-029}.  The electron, muon, jet and \metvec
systematic uncertainties described above are found to have a very small impact.

Theoretical cross-section uncertainties are taken into account for all
backgrounds estimated using simulation.  For \DYjets production, uncertainties
are taken from Ref.~\cite{EXOT-2015-07} and include variations of the PDF
sets, scale, \alphas, beam energy, electroweak corrections and photon-induced
corrections. A single 90\% CL eigenvector variation uncertainty is used, based
on the CT14nnlo PDF set. The variations amount to a $\sim$$5\%$ uncertainty in
the total number of \DYjets events within the acceptance. For \diboson production,
an uncertainty of $10\%$ is used~\cite{Campbell:2011bn,Gleisberg:2008ta}. For
\ttbar~\cite{Czakon:2011xx} and \singletop~\cite{Aliev:2010zk, Kant:2014oha}
production, the assigned $6\%$ uncertainty is based on PDF, scale and top-quark
mass variations.  Additional uncertainties related to initial- and final-state
radiation modelling, tune and (for \ttbar only) the choice of 
\texttt{hdamp} parameter value in \POWHEGBOXV{ v2}, which controls the amount
of radiation produced by the parton shower,  are also taken into
account~\cite{ATL-PHYS-PUB-2016-004}.  The uncertainty due to the hadronisation model is
evaluated by comparing \ttbar events generated with \POWHEGBOXV{ v2} interfaced
to either \HERWIGpp~\cite{Herwigpp} or \PYTHIAV{6}. To estimate the uncertainty
in generating the hard scatter, the \POWHEG{} and MG5\_aMC@NLO event generators are
compared, both interfaced to the \HERWIGpp  parton shower model. The
uncertainties in the \Wjets cross section have a negligible impact in the
\hadhad channel and the \Wjets simulation is not used in the \lephad channel.

For MSSM Higgs boson samples, various sources of uncertainty 
which impact the signal acceptance are considered. The impact from varying the factorisation
and renormalisation scales up and down by a factor of two, either coherently or
oppositely, is taken into account.  Uncertainties due to the modelling of
initial- and final-state radiation, as well as multiple parton interaction are
also taken into account. These uncertainties are estimated from variations of
the \PYTHIAV{8} A14 tune~\cite{ATL-PHYS-PUB-2014-021} for the  $b$-associated
production and  the AZNLO \PYTHIAV{8} tune~\cite{AZNLO:2014} for the gluon--gluon
fusion production.  The envelope of the variations resulting from the use of
the alternative PDFs in the
PDF4LHC15$\_$nlo$\_$nf4$\_$30\,(PDF4LHC15$\_$nlo$\_$100)~\cite{Ball:2014uwa}
set is used to estimate the PDF uncertainty for the $b$-associated
(gluon--gluon fusion) production.  The total uncertainty for the MSSM Higgs boson
samples is typically 1--4\%, which is dominated by variations of the radiation
and multiple parton interactions, with minor impact from scale variations. 
The \Zprime signal acceptance uncertainties are expected to be negligible.

For both the MSSM Higgs boson and \Zprime samples, uncertainties in the
integrated cross section are not included in the fitting procedure used to
extract experimental cross-section limits. The uncertainty for \Zprime is
included when overlaying model cross sections, in which case it is calculated
using the same procedure as for the \DYjets background.

\subsection{Uncertainties in data-driven estimates}
\label{sec:sys:data}

Uncertainties in the \multijet estimate for the \hadhad channel
(Section~\ref{sec:hadhad:multijet}) arise from the fake-factors \fDJ.
These include a 10--50\% uncertainty from the limited size of the \djcr and an
uncertainty of up to 50\% from the subtraction of the non-\multijet
contamination.  An additional uncertainty is considered when applying the
fake-factors in the \btag category, which accounts for changes in the jet
composition with respect to the inclusive selection of the \djcr. As the
differences are extracted from comparisons in control regions, they are
one-sided.

The uncertainty in the fake-rates used to weight simulated non-\multijet 
events in the \hadhad channel (Section~\ref{sec:hadhad:non-multijet})
is dominated by the limited size of the fakes regions and can reach 40\%.

Uncertainties in the \multijet estimate for the \lephad channel 
(Section~\ref{sec:lephad:multijet}) arise from the fake-factors \fMJ
and \flep.  The applicability of \fMJ measured in \mjcr to \failcr is
investigated by studying \fMJ as a function of the lepton isolation and the
observed differences are assigned as a systematic uncertainty. The statistical
uncertainty from the limited size of \mjcr is significant, particularly for the
smaller 2015 dataset. The impact of a potential mismodelling in the subtraction
of simulated non-\multijet events containing non-isolated leptons is
investigated by varying the subtraction by 50\%, but is found to be small
compared to the other sources of systematic uncertainty.  A constant
uncertainty of 20\% in \fMJ is used to envelop these variations. A 50\%
uncertainty is assigned to the sequential
$|\Delta\phi(\ptvec^{\tauhadvis},\metvec)|$ correction.

The applicability of \flep measured in \flcr to events in \mjcr is investigated
by altering the $\mT(\ptvec^{\ell},\metvec)$ selection and the observed
differences are assigned as a systematic uncertainty.  A 20\% uncertainty in
the background subtraction in \flcr is considered, motivated by observations of
the tau identification performance in \Wjets events.  The statistical
uncertainty from the limited size of \flcr is also considered, but is
relatively small. The total uncertainty in \flep is 5--50\%.

Uncertainties in the data-driven \Wjets and \ttbar estimates for the \lephad
channel (Section~\ref{sec:lephad:non-multijet}) arise from the fake-factors \fW
and the subtraction of contributions from \failcr.  The applicability of \fW
measured in \wcr to \failcr is investigated by studying \fW as a function
of $\mT(\ptvec^{\ell},\metvec)$ and the observed differences (up to $\sim$10\%)
are assigned as a systematic uncertainty. A 30\% uncertainty is assigned to the
sequential $|\Delta\phi(\ptvec^{\tauhadvis},\metvec)|$ correction, based on
variations observed as a function of \tauhadvis \pt.  Due to the large
contamination for \btag events in \wcr, a 50\% uncertainty is assigned to
the correction factor applied to the \bveto parameterisation.  The subtraction
of the simulated samples in \failcr is affected by experimental uncertainties
and uncertainties in production cross sections, which amount to 10\%. The total
uncertainty in the \multijet estimate in \failcr is also propagated to the
subtraction.

\section{Results}
\label{sec:results}

\begin{figure}[tp]
  \centering
  \subfigure[\lephad \bveto category\label{fig:mttot1}]{\includegraphics[width=0.45\textwidth]{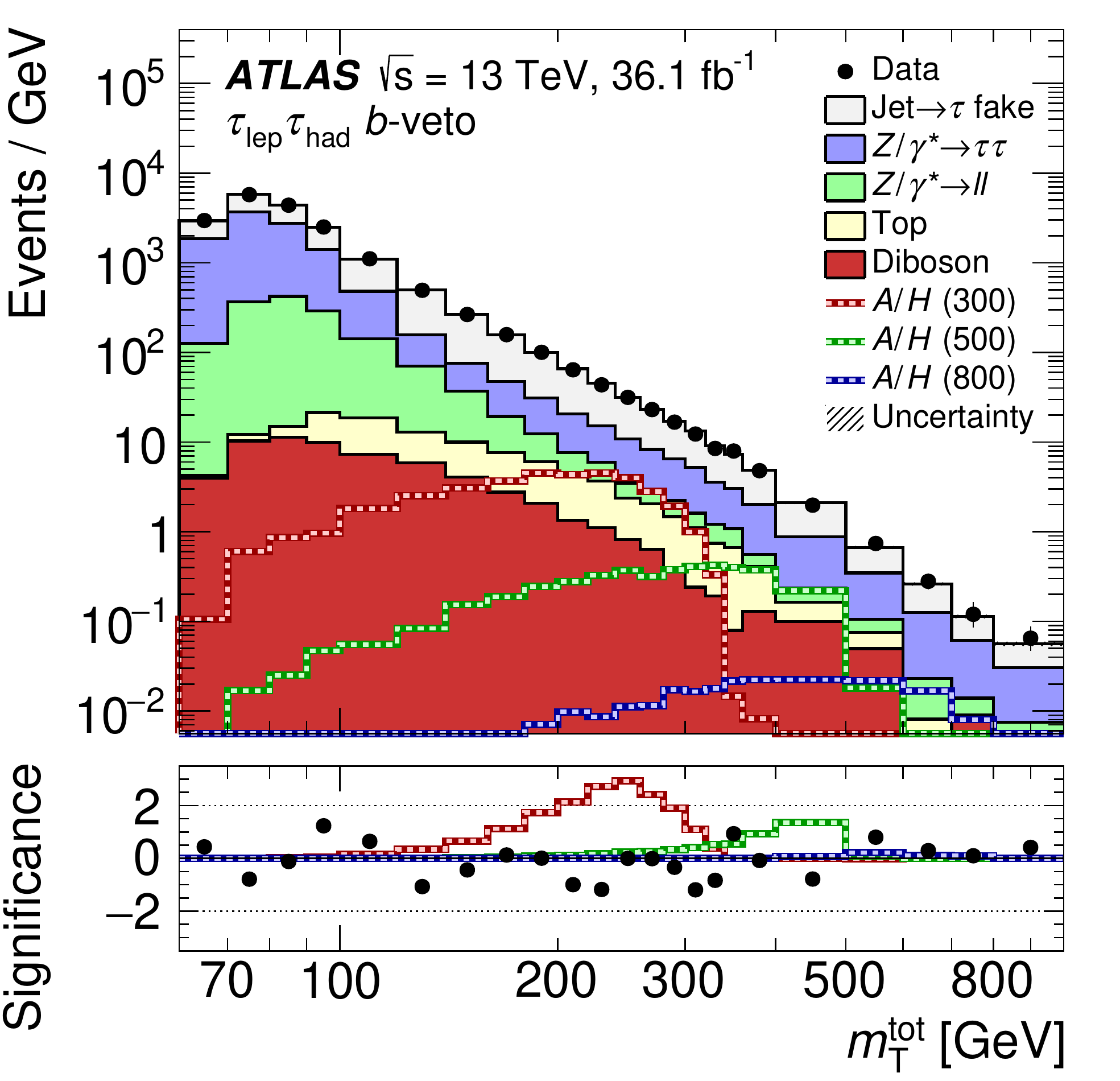}}
  \subfigure[\lephad \btag category\label{fig:mttot2}]{\includegraphics[width=0.45\textwidth]{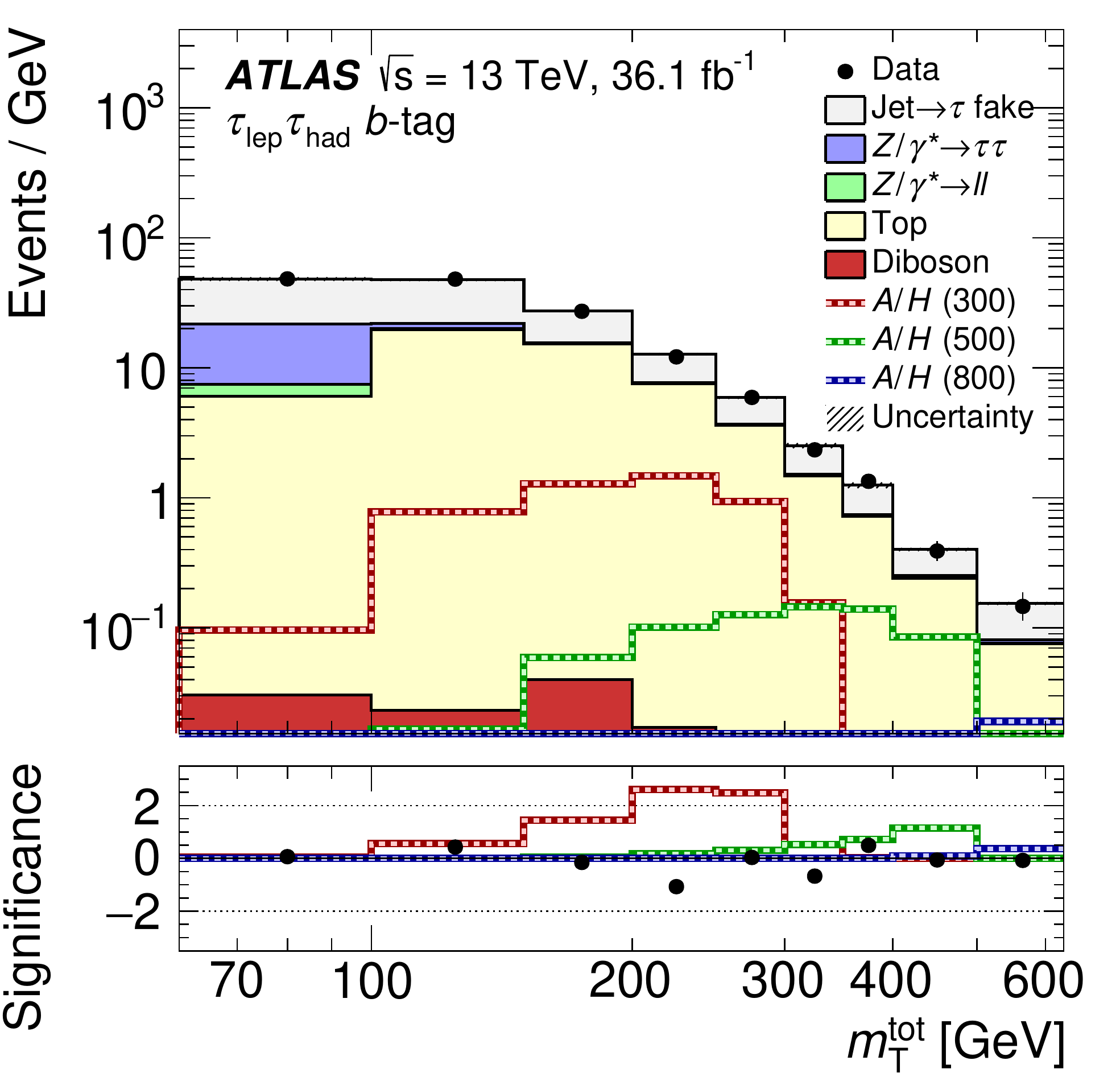}}\\
  \subfigure[\hadhad \bveto category\label{fig:mttot3}]{\includegraphics[width=0.45\textwidth]{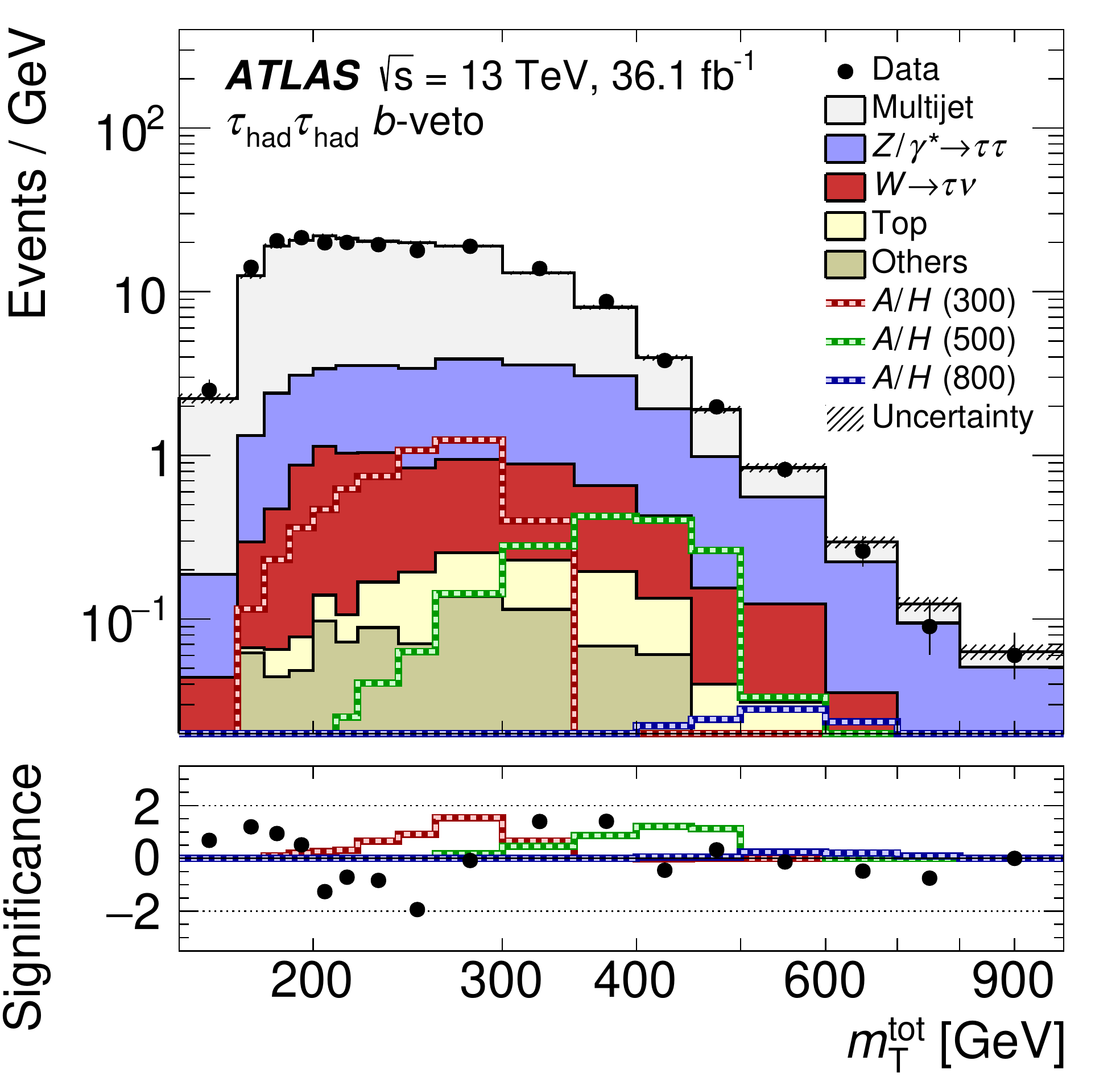}}
  \subfigure[\hadhad \btag  category\label{fig:mttot4}]{\includegraphics[width=0.45\textwidth]{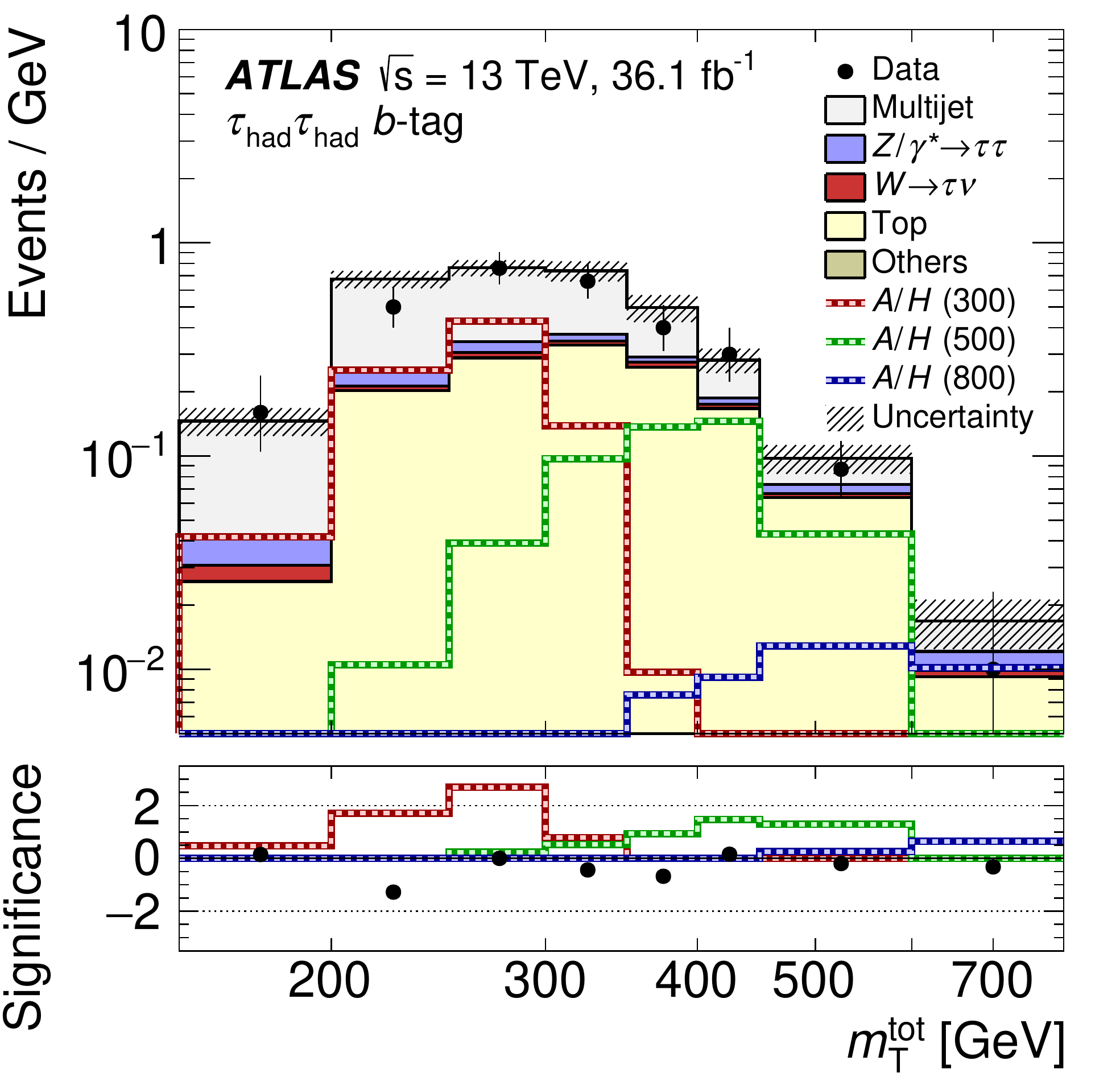}}
  \caption{
    Distributions of $\mttot$ for the (a) \bveto and (b) \btag categories of the
    \lephad channel and the (c) \bveto and (d) \btag categories of the \hadhad
    channel.  The label  ``\Others'' refers to contributions from \diboson, \DYlljets
    and \Wlnujets production. In the \lephad channel, events containing \tauhadvis candidates that
    originate from jets are removed from all processes other than \jetfakes.  The
    binning displayed is that entering into the statistical fit discussed in
    Section~\ref{sec:results}, with minor modifications needed to combine the 
    \lephad channels and with underflows and overflows included in the first
    and last bins, respectively. The predictions and uncertainties for the
    background processes are obtained from the fit under the hypothesis of no
    signal.  The combined prediction for $A$ and $H$ bosons with masses of 300,
    500 and \SI{800}{\GeV} and $\tan\beta=10$ in the \hmssm scenario are
    superimposed. The significance of the data given the fitted model and 
    its uncertainty is computed in each bin following Ref.~\cite{Diego:sig}
    and is shown in the lower panels. The expected significance of the hypothetical 
    Higgs boson signals are also overlaid.
}
\label{fig:mttotPF}
\end{figure}

\begin{figure}[tp]
\begin{center}
\includegraphics[width=0.45\columnwidth]{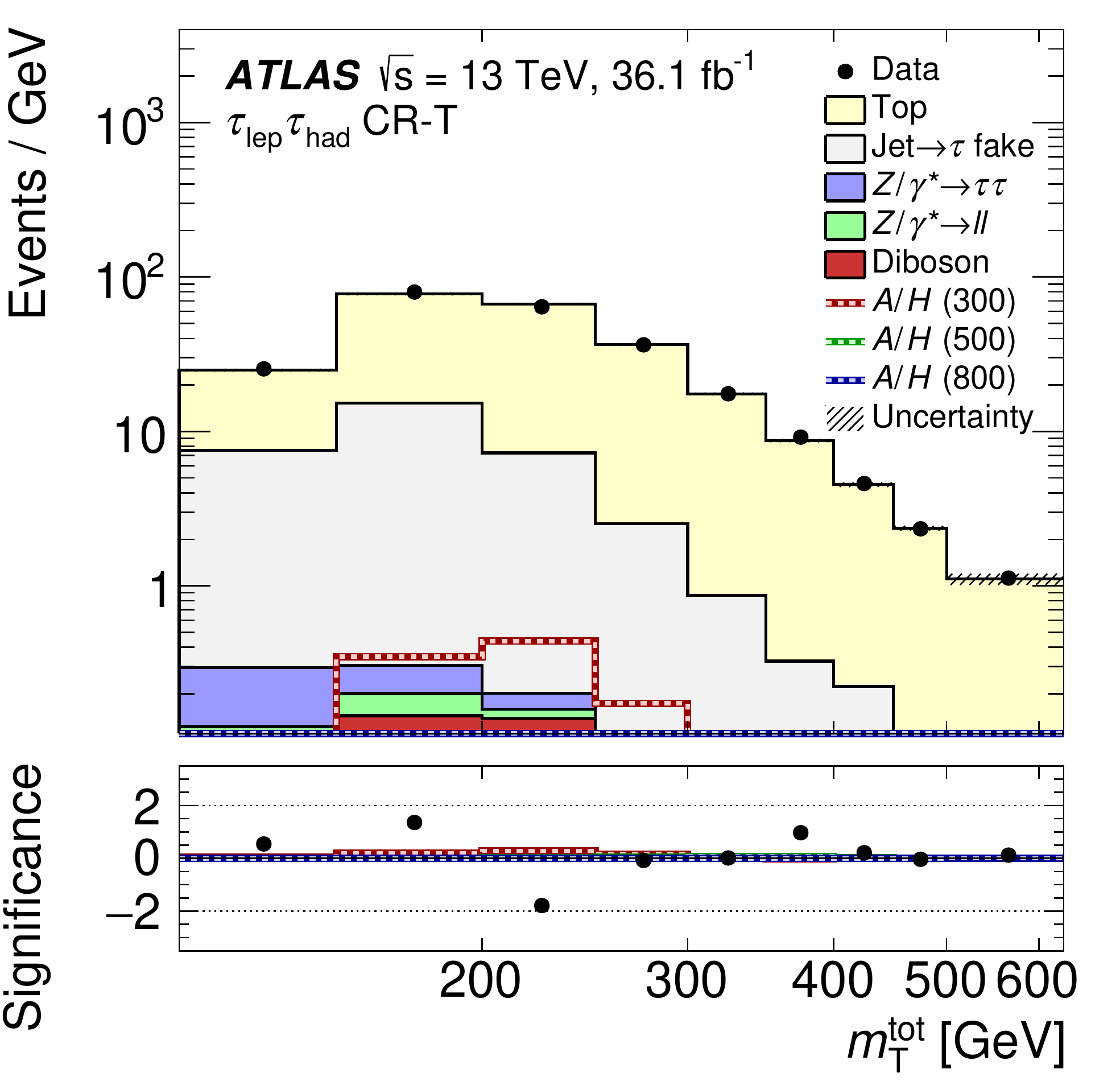}
\caption{%
    Distribution  of $\mttot$ in the \ttbar enriched control region (CR-T) 
    of the \lephad channel.
    Events containing \tauhadvis candidates that
    originate from jets are removed from all processes other than \jetfakes.  The
    binning displayed is that entering into the statistical fit discussed in
    Section~\ref{sec:results}, with minor modifications needed to combine the
    \lephad channels and with underflows and overflows included in the first
    and last bins, respectively. The predictions and uncertainties for the
    background processes are obtained from the fit under the hypothesis of no
    signal.  The combined prediction for $A$ and $H$ bosons with masses of 300,
    500 and \SI{800}{\GeV} and $\tan\beta=10$ in the \hmssm scenario are
    superimposed. The significance of the data given the fitted model and 
    its uncertainty is computed in each bin following Ref.~\cite{Diego:sig}
    and is shown in the lower panel. The expected significance of the hypothetical 
    Higgs boson signals are also overlaid.
}

\label{fig:tcr}
\end{center}
\end{figure}

The number of observed events in the signal regions of the \lephad and \hadhad
channels together with the predicted event yields from signal and background
processes are shown in Table~\ref{tab:data-pred-comparison}. In the \lephad
channel, all events estimated using the data-driven fake-factor technique are
grouped as \jetfakes, while events where the \tauhadvis originates from a jet
are removed from the other processes. In the \hadhad channel, the \multijet
process is estimated using the fake-factor technique while contributions from
all other processes are estimated using simulation with data-driven corrections
for the \tauhadvis candidates that originate from jets.  The numbers are given
before (pre-fit) and after (post-fit) applying the statistical fitting
procedure described in Section~\ref{sec:fitmodel}.  The observed event yields
are compatible with the expected event yields from SM processes, within
uncertainties.  The $\mttot$ distributions in the signal regions are shown in
Figures~\ref{fig:mttot1}--\ref{fig:mttot4} and in the CR-T in
Figure~\ref{fig:tcr}.

\begin{table}[tp]
\footnotesize

\begin{center}

\caption{
Observed number of events and predictions of signal and background
contributions in the \bveto and \btag categories of the \lephad and \hadhad
channels.  The background predictions and uncertainties (including both the
statistical and systematic components) are obtained before (pre-fit) and after
(post-fit) applying the statistical fitting procedure discussed in
Section~\ref{sec:results}.  The individual uncertainties are correlated, and do
not necessarily add in quadrature to the total background uncertainty. The
label ``\Others'' refers to contributions from \diboson, \DYlljets and
\Wlnujets production.  In the \lephad channel, events containing a \tauhadvis
candidate that originate from jets are removed from all processes other than
\jetfakes. The expected pre-fit contributions from $A$ and $H$ bosons with
masses of 300, 500 and \SI{800}{\GeV} and $\tan\beta=10$ in the \hmssm scenario
are also shown.
} 
\label{tab:data-pred-comparison}

  \begin{tabular}{ll r@{$\,\pm\,$}r r@{$\,\pm\,$}r r@{$\,\pm\,$}r r@{$\,\pm\,$}r }
\toprule
&         &  \multicolumn{4}{c}{\bveto}  &  \multicolumn{4}{c}{\btag} \\
Channel & Process  &  \multicolumn{2}{c}{pre-fit}  &  \multicolumn{2}{c}{post-fit} 
         &  \multicolumn{2}{c}{pre-fit}  &  \multicolumn{2}{c}{post-fit} \\
\midrule 
\lephad & \DYtt         & \num{92000}  & \num{11000} & \num{96400} & \num{1600} & \num{670}   & \num{140} & \num{690}  & \num{70}   \\
&\Diboson               & \num{880}    & \num{100}   & \num{920}   & \num{70}   & \num{6.3}   & \num{1.7} & \num{6.5}  & \num{1.4}  \\
&\ttbar and \singletop  & \num{1050}   & \num{170}   & \num{1090}  & \num{130}  & \num{2800}  & \num{400} & \num{2680} & \num{80}   \\
&\jetfakes              & \num{83000}  & \num{5000}  & \num{88800} & \num{1700} & \num{3000}  & \num{400} & \num{3390} & \num{170}  \\
&\DYll                  & \num{15800}  & \num{1200}  & \num{16200} & \num{700}  & \num{86}    & \num{21}  & \num{89}   & \num{16}   \\
\midrule                                                                                                                          
&SM Total               & \num{193000} & \num{13000} & \num{203400}& \num{1200} & \num{6500}  & \num{600} & \num{6850} & \num{120}  \\
&Data                   & \multicolumn{4}{c}{\num{203365}}  & \multicolumn{4}{c}{\num{6843}} \\ 
\midrule                                                                                                         
&$A/H$ (300)            & \num{720}    &\num{80}     & \multicolumn{2}{c}{--}   & \num{236}   & \num{32}    & \multicolumn{2}{c}{--}   \\
&$A/H$ (500)            & \num{112}    &\num{11}     & \multicolumn{2}{c}{--}   & \num{39}    & \num{5}     & \multicolumn{2}{c}{--}   \\
&$A/H$ (800)            & \num{10.7}   &\num{1.1}    & \multicolumn{2}{c}{--}   & \num{4.8}   & \num{0.6}   & \multicolumn{2}{c}{--}   \\
\midrule 
\hadhad & \Multijet    & \num{3040} & \num{240}  & \num{3040} & \num{90}  & \num{106} & \num{32}  & \num{85}   & \num{10}   \\
&\DYtt                 & \num{610}  & \num{230}  & \num{770}  & \num{80}  & \num{7.5} & \num{2.9} & \num{8.6}  & \num{1.3}  \\
&\Wtaunujets           & \num{178}  & \num{31}   & \num{182}  & \num{15}  & \num{4.0} & \num{1.0} & \num{4.1}  & \num{0.5}  \\
&\ttbar and \singletop & \num{26}   & \num{9}    & \num{29}   & \num{4}   & \num{60}  & \num{50}  & \num{74}   & \num{15}   \\
&\Others               & \num{25}   & \num{6}    & \num{27.4} & \num{2.1} & \num{1.0} & \num{0.5} & \num{1.1} & \num{0.4} \\
\midrule                                                                                                                  
&SM Total              & \num{3900} & \num{400}  & \num{4050} & \num{70}  & \num{180} & \num{60}  & \num{173}  & \num{16}   \\
&Data                  & \multicolumn{4}{c}{\num{4059}}     & \multicolumn{4}{c}{\num{154}} \\
\midrule                                                                                      
&$A/H$ (300)           & \num{130}   & \num{50}    & \multicolumn{2}{c}{--}   & \num{44}  & \num{19}   & \multicolumn{2}{c}{--} \\
&$A/H$ (500)           & \num{80}    & \num{33}    & \multicolumn{2}{c}{--}   & \num{28}  & \num{12}   & \multicolumn{2}{c}{--} \\
&$A/H$ (800)           & \num{11}    & \num{4}     & \multicolumn{2}{c}{--}   & \num{5.1} & \num{2.2}  & \multicolumn{2}{c}{--} \\
\bottomrule
\end{tabular}

\end{center}
\end{table}

\subsection{Fit model}
\label{sec:fitmodel}

The parameter of interest is the signal strength, $\mu$.  It is defined as the
ratio of the observed to the predicted value of the cross section times
branching fraction, where the prediction is evaluated at a particular
point in the parameter space of the theoretical model in question (MSSM or \Zprime
benchmark scenarios).  Hence, the value $\mu=0$ corresponds to the absence of a
signal, whereas the value $\mu=1$ indicates the presence of a signal as
predicted by the model.  To estimate $\mu$, a likelihood function constructed
as the product of Poisson probability terms is used. A term is included for
each bin in the \mttot distributions from the \ehad, \muhad and \hadhad
channels. When fitting MSSM models to the data, the distributions are separated
into \btag and \bveto events to enhance sensitivity to the gluon--gluon fusion and
$b$-associated production modes, while the inclusive distributions are used for
\Zprime models. In all cases, the distributions in the CR-T regions of
the \ehad and \muhad channels are added, which help constrain uncertainties in
the \ttbar background.  Signal and background predictions depend on systematic
uncertainties, which are parameterised as nuisance parameters that are
constrained using Gaussian probability density functions.   The asymptotic
approximation is used with the test statistic
$\tilde{q}_\mu$~\cite{Cowan:2010js} to compare the likelihoods 
of the null hypothesis (SM only) and the assumed signal hypothesis (SM
plus signal) given the data. The bin widths are chosen to ensure a sufficient number of
background events in each bin.  The results from the \ehad, \muhad and \hadhad
channels are combined to improve the sensitivity to signal. For ditau resonance 
masses below about \SI{0.6}{\TeV}, the sensitivity is dominated by the \lephad channels, 
while the \hadhad channel is most sensitive in the higher mass range.

\subsection{Cross-section limits}

\begin{figure}[tp]
  \centering
  \subfigure[$\phi\to\tau\tau$ (gluon--gluon fusion production)\label{fig:limits-xsec1}]{\includegraphics[width=0.45\textwidth]{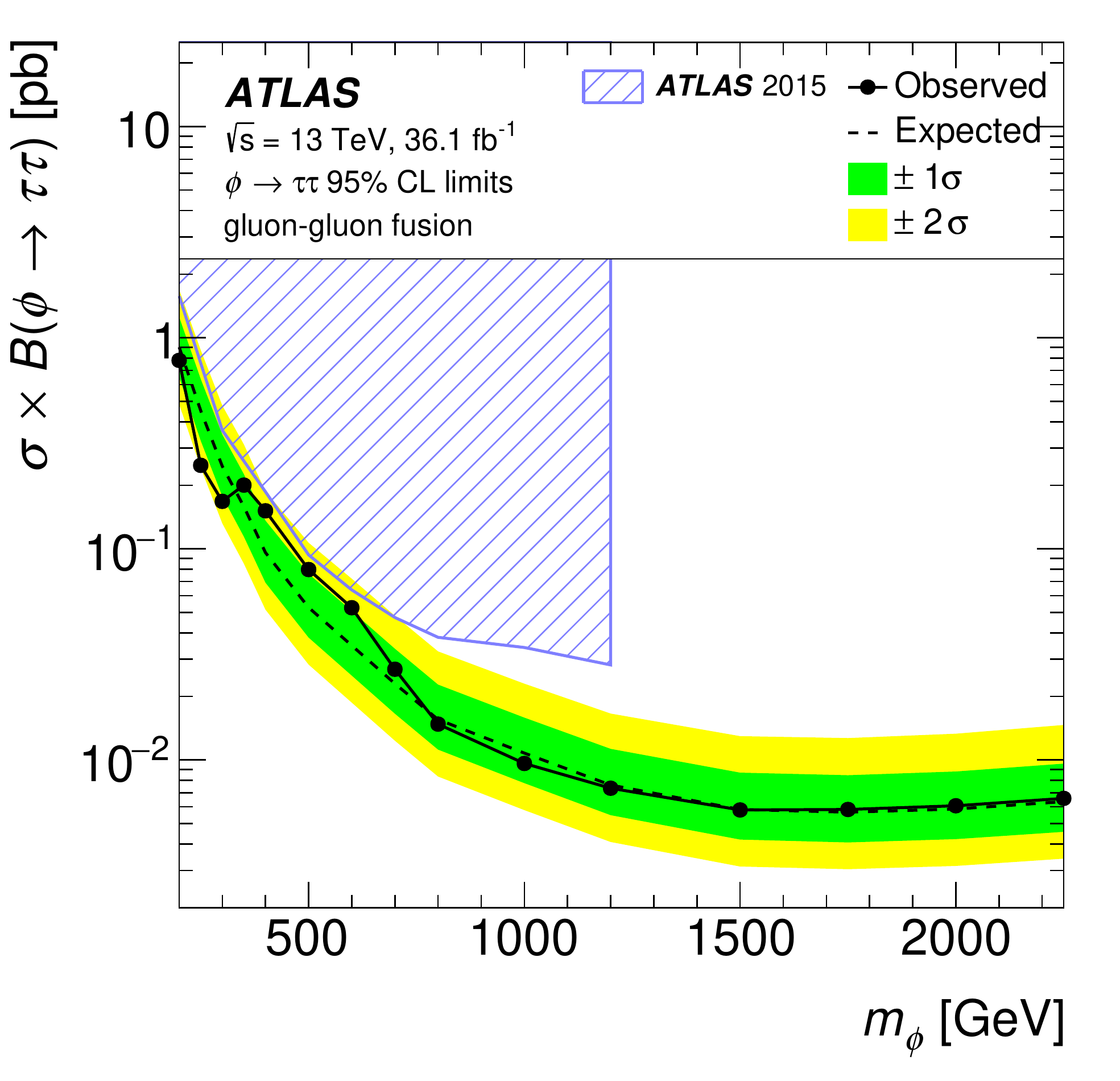} }
  \subfigure[$\phi\to\tau\tau$ ($b$-associated production)]{\includegraphics[width=0.45\textwidth]{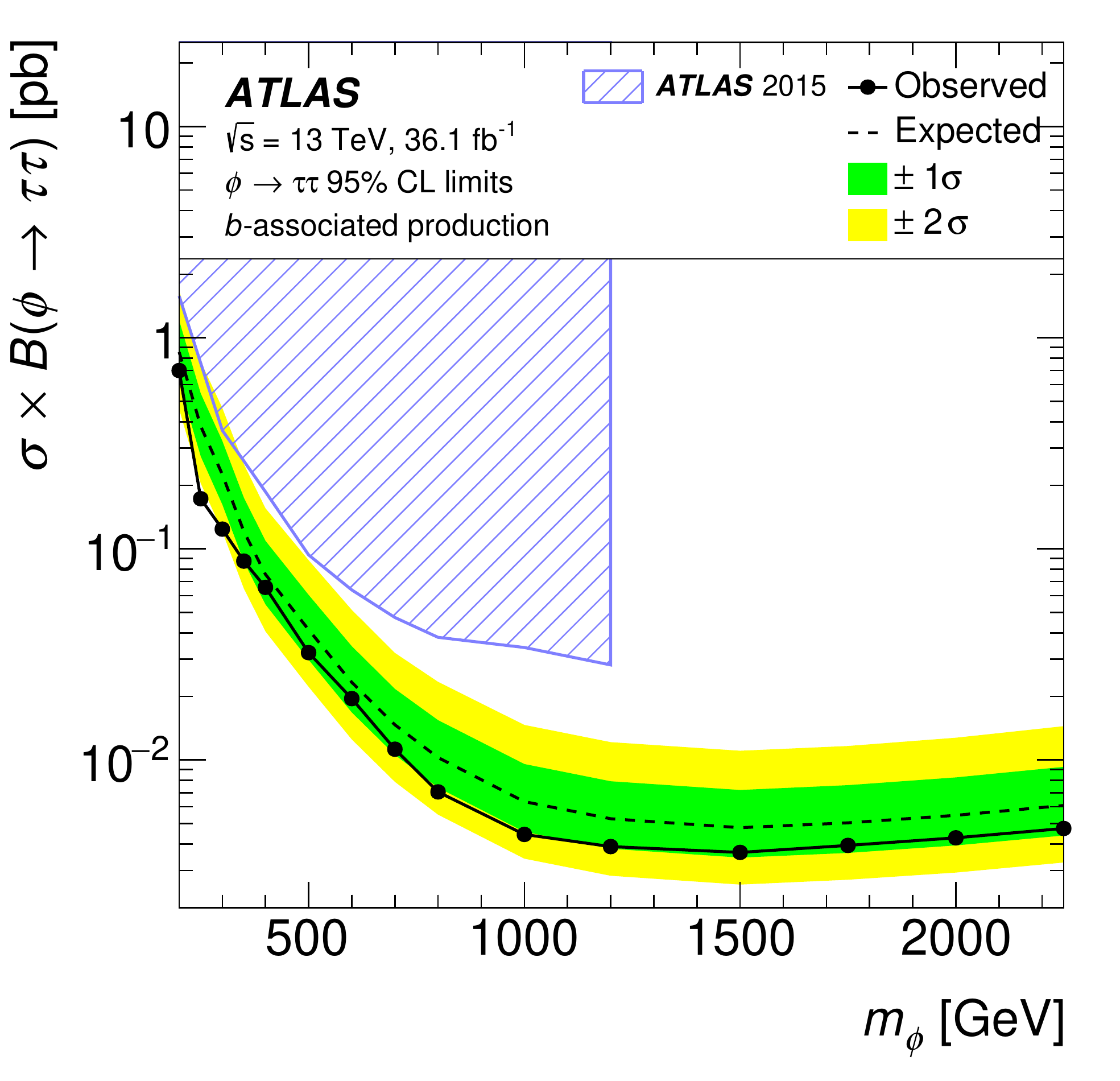} }\\
  \subfigure[$\Zprime\to\tau\tau$\label{fig:limits-xsec3}]{\includegraphics[width=0.45\textwidth]{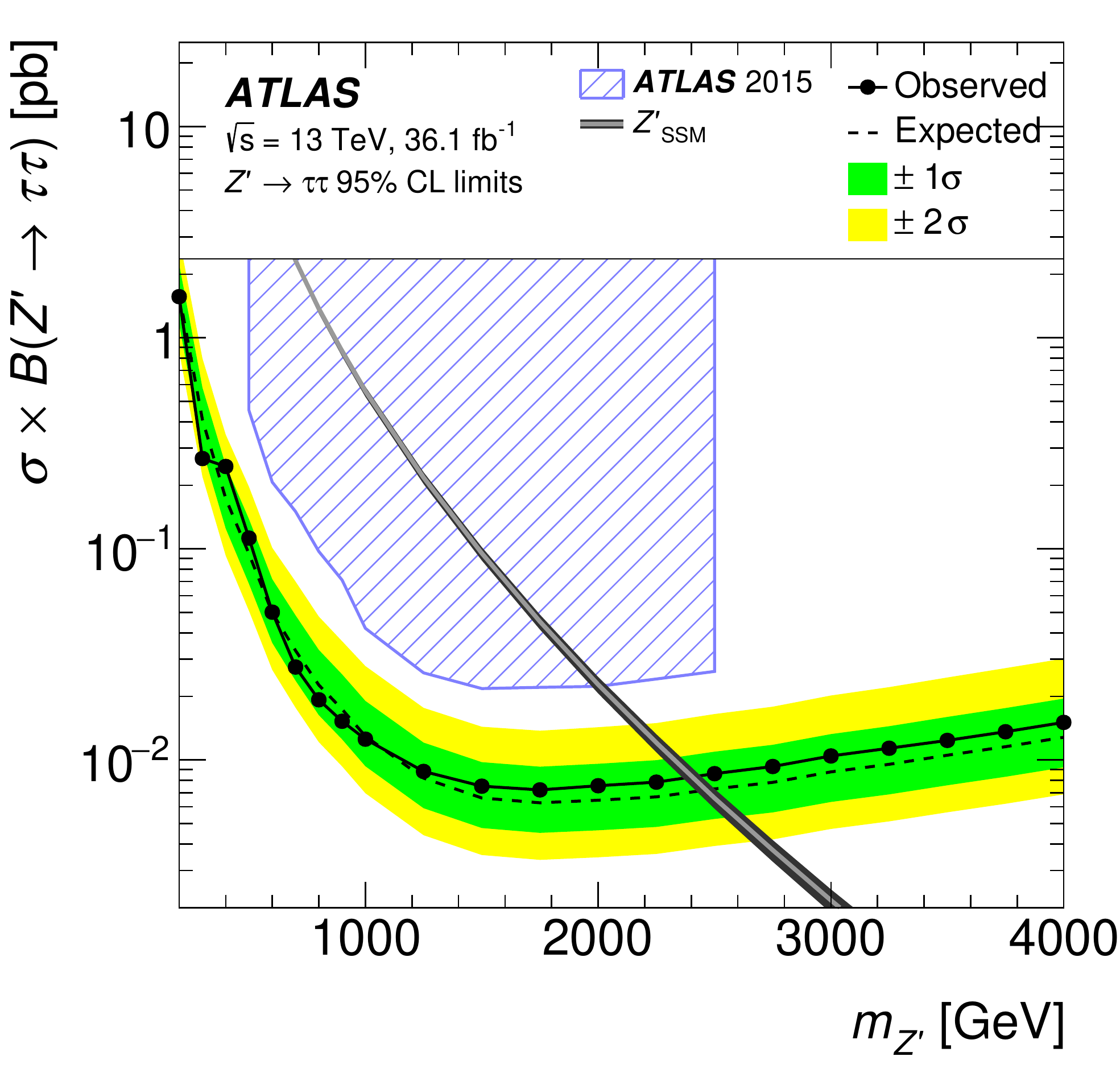} }
  
  \caption{
    The observed and expected 95\% CL upper limits on the production 
    cross section times branching fraction for a scalar 
    boson produced via (a) gluon--gluon fusion and (b) $b$-associated production,
    and for (c) gauge bosons. The limits are calculated 
    from a statistical combination of the \lephad and \hadhad channels. 
    The excluded regions from the 2015 data ATLAS search~\cite{HIGG-2015-08}
    are depicted by the hatched blue fill.
    The predicted cross section for a \ZprimeSSM boson is overlaid 
    in (c), where the band depicts the total uncertainty.}
\label{fig:limits-xsec}
\end{figure}

\begin{figure}[tp]
  \centering
  \subfigure[Observed\label{fig:limits-bb-vs-gg:obs}]{\includegraphics[width=0.45\textwidth]{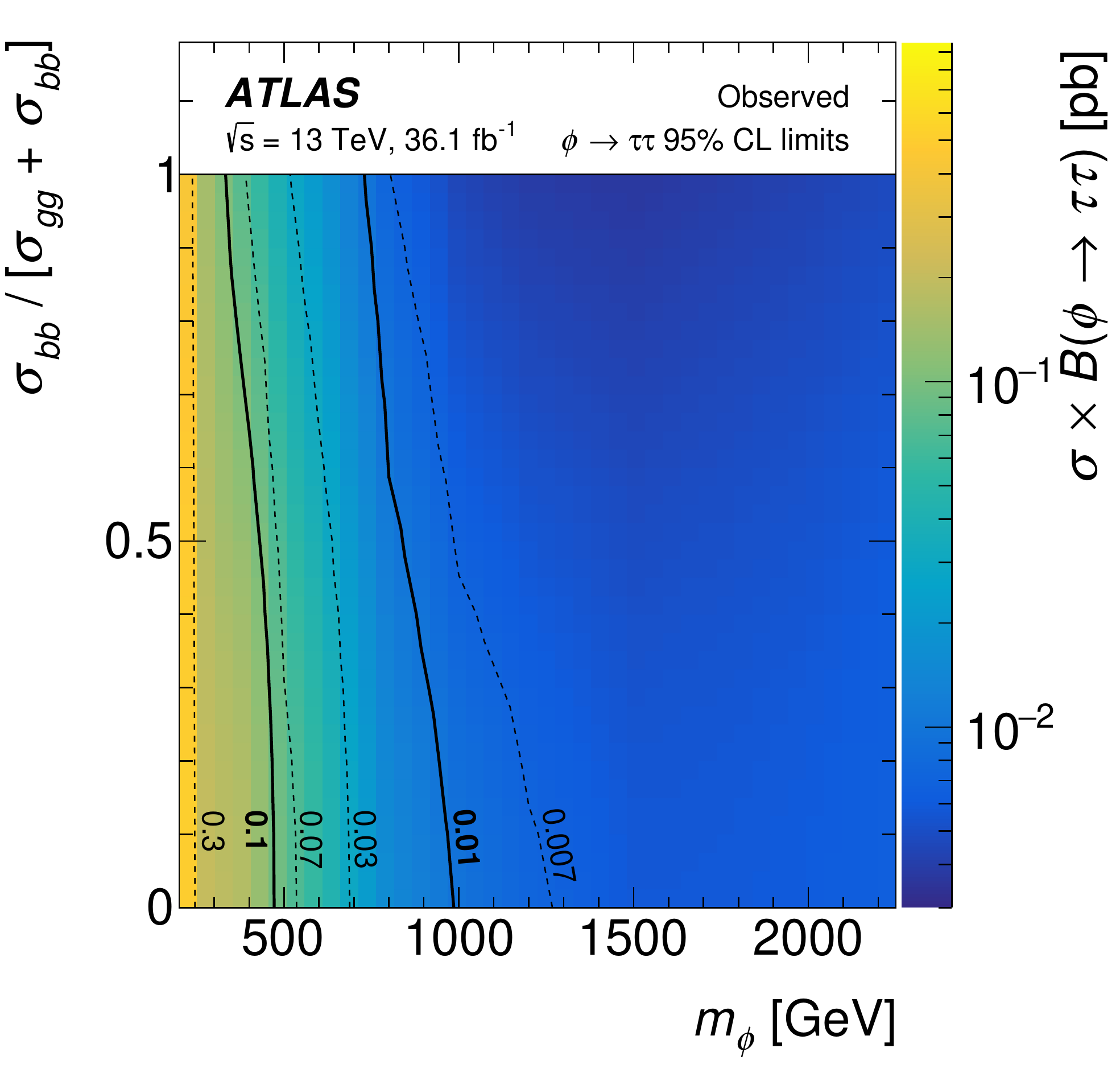}}
  \hspace{2em}
  \subfigure[Expected\label{fig:limits-bb-vs-gg:exp}]{\includegraphics[width=0.45\textwidth]{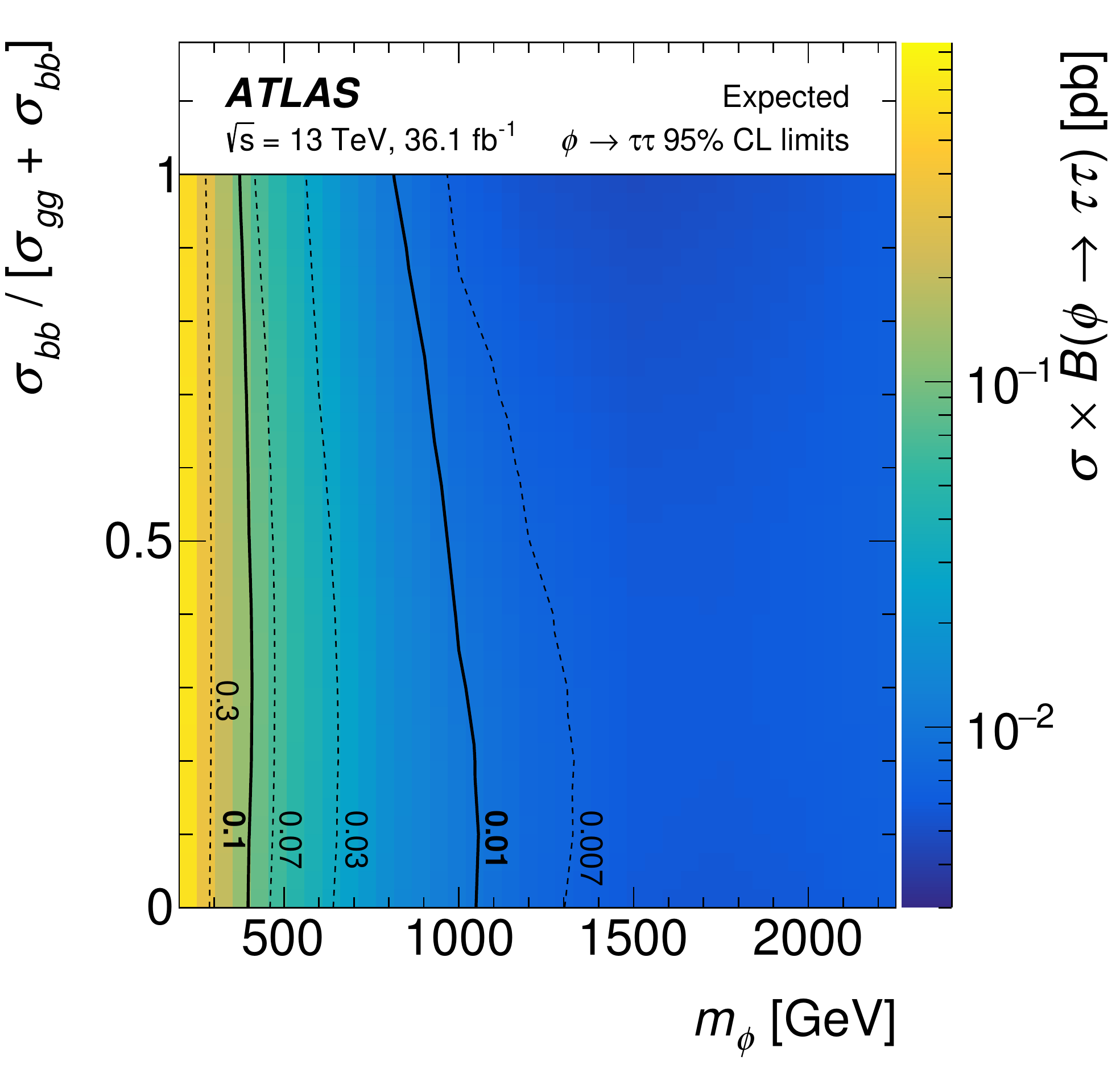}}
  \caption{
    The (a) observed and (b) expected 95\% CL upper limit on the production cross section times
    branching fraction for $\phi\to\tau\tau$ as a function of the fractional 
    contribution from $b$-associated production and the scalar boson mass. 
    The solid and dashed lines represent contours of fixed $\sigma\times B$.}
\label{fig:limits-bb-vs-gg}
\end{figure}

The data are found to be in good agreement with the predicted background
yields, and the results are given in terms of exclusion limits.  These are set
using the modified frequentist CL$_\mathrm{s}$ method~\cite{CLs_2002}.  Upper
limits on the cross section times branching fraction for $\phi$ and \Zprime
bosons are set at the 95\% confidence level (CL) as a function of the boson
mass. They are obtained by multiplying the extracted limits on $\mu$ by the
respective predicted cross sections. The $\phi$ boson limits assume the natural
width of the boson to be negligible compared to the experimental resolution (as
expected over the probed MSSM parameter space). They cover the mass range
\hmin--\SI{\hmax}{\TeV} and are shown separately for gluon--gluon fusion and
$b$-quark associated production.  The limits on \Zprime bosons are calculated
assuming an SSM \Zprime and extend up to \SI{\zmax}{\TeV}. The limits are shown
in Figures~\ref{fig:limits-xsec1}--\ref{fig:limits-xsec3}.  They are in the
range \limgghigh--\SI{\limgglow}{\pb} (\limbbhigh--\SI{\limbblow}{\pb}) for
gluon--gluon fusion ($b$-associated) production of scalar bosons with masses of
\hmin--\SI{\hmax}{\TeV} and \limzphigh--\SI{\limzplow}{\pb} for Drell--Yan
production of \Zprime bosons with masses of \zmin--\SI{\zmax}{\TeV}.  A small
downward fluctuation at a mass of $\sim$\SI{0.3}{\TeV} is observed in all
limits, while a small upward fluctuation for  gluon--gluon fusion and \Zprime
bosons is seen around \SI{0.5}{\TeV} and a broad deficit is seen for the
$b$-quark associated production over the entire mass range. These features
arise primarily because of a deficit of events in the range 200--\SI{250}{\GeV}
followed by a mild excess in the range 300--\SI{400}{\GeV} in
Figure~\ref{fig:mttot3}, and by a consistent deficit of events across the whole
range in Figure~\ref{fig:mttot4}.  Modifications of the \Zprime chiral coupling
structure can result in changes of up to 40\% in the \Zprime cross-section
limits. Reducing the \Zprime width can improve the limits by up to $\sim$30\%,
while increasing the width to 36\% can degrade the limits by up to $\sim$70\%.
Figures~\ref{fig:limits-bb-vs-gg:obs} and \ref{fig:limits-bb-vs-gg:exp} show
the observed and expected 95\% CL upper limits on the production cross section
times branching fraction for $\phi\to\tau\tau$ as a function of the fractional
contribution from $b$-associated production ($\sigma_{bb} / [\sigma_{bb} +
\sigma_{gg}]$) and the scalar boson mass.

\begin{figure}[tp]
  \centering
  \subfigure[gluon--gluon fusion production\label{fig:sys:gg}]{\includegraphics[width=0.45\textwidth]{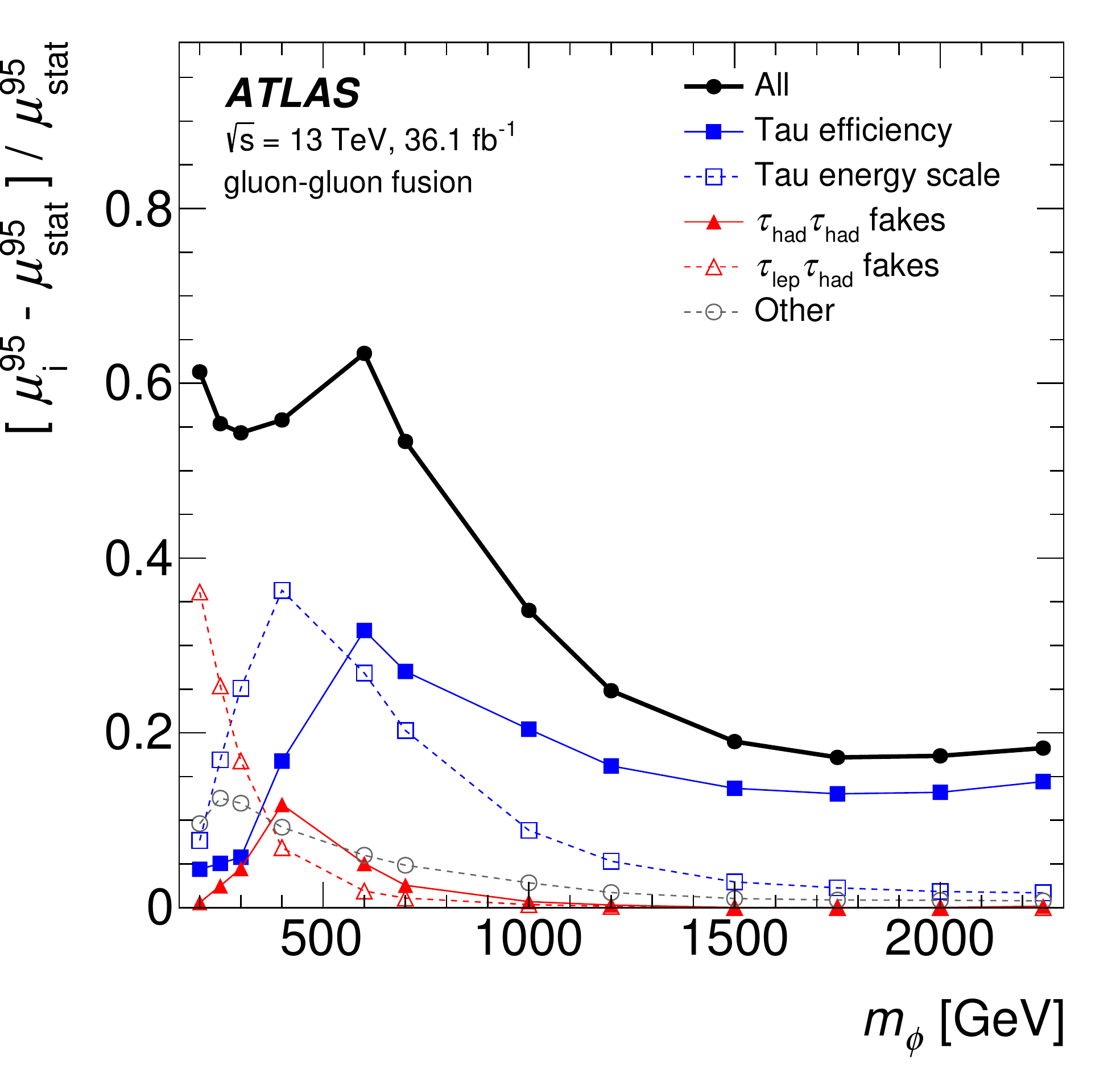}}
  \hspace{2em}
  \subfigure[$b$-associated production\label{fig:sys:bb}]{\includegraphics[width=0.45\textwidth]{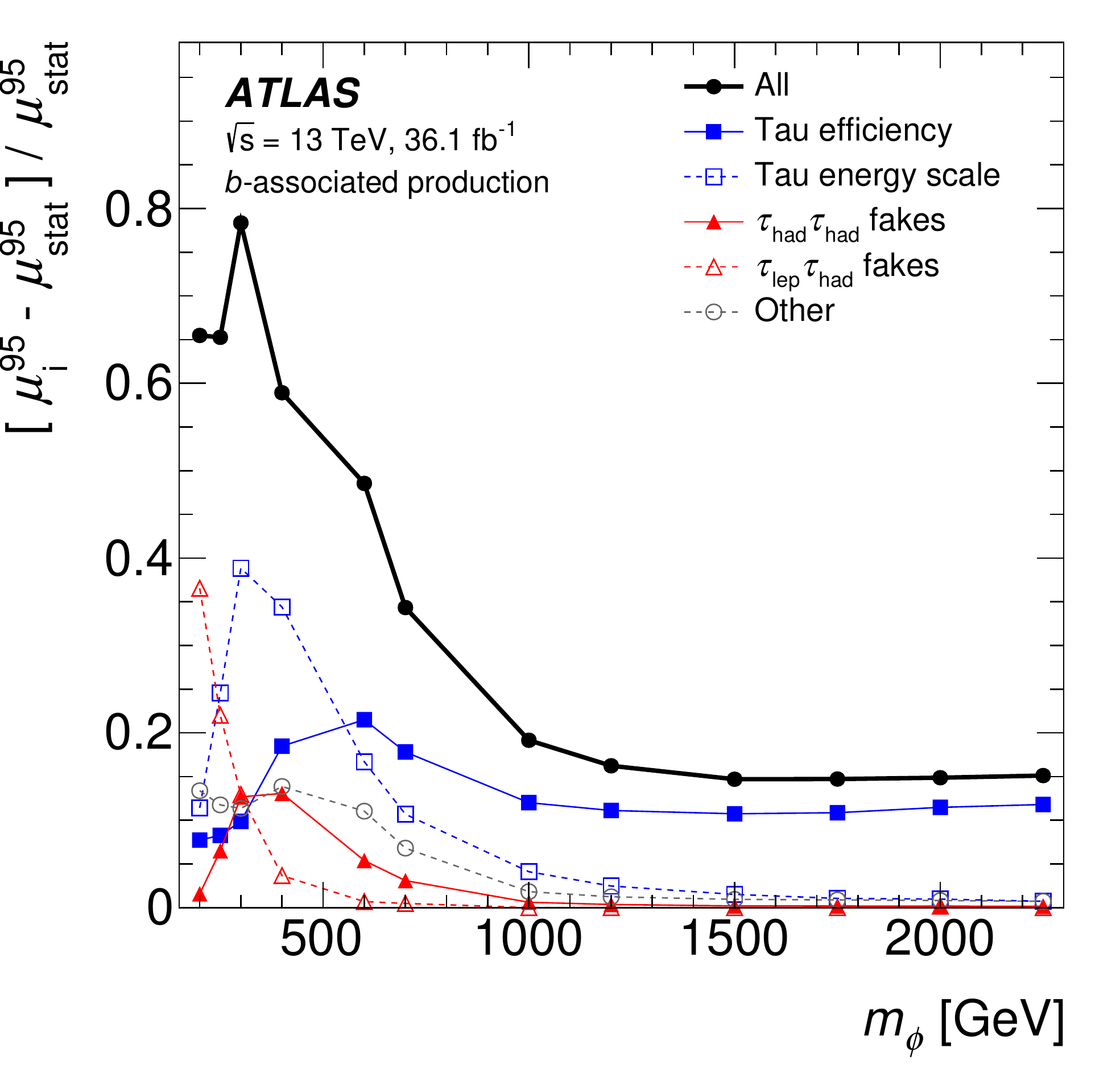}}
  
  \caption{Impact of major groups of systematic uncertainties on the 
    $\phi\to\tau\tau$ 95\% CL cross section upper limits as a function 
    of the scalar boson mass, separately for the (a) gluon--gluon fusion 
    and (b) $b$-associated production mechanisms. 
  }
\label{fig:systematics}
\end{figure}

The impact of systematic uncertainties on the $\phi\to\tau\tau$ 95\% CL cross
section upper limits are calculated by comparing the expected upper limit in
the case of no systematic uncertainties, $\mu^{95}_\textrm{stat}$, with a limit
calculated by introducing a group of systematic uncertainties, $\mu^{95}_{i}$.
The systematic uncertainty impacts are shown in Figure~\ref{fig:sys:gg} for
gluon--gluon fusion production and Figure~\ref{fig:sys:bb} for $b$-associated
production as functions of the scalar boson mass.  The major uncertainties are
grouped according to their origin, while minor uncertainties are collected as
``Others''. 

In the low mass range, the sensitivity is dominated by the \lephad channel, and
the major uncertainties arise from the estimate  of the dominant \Wjets
background. Due to the large contribution the fit is able to significantly
constrain the uncertainties in this background.  In the intermediate mass range
the tau energy scale uncertainty becomes dominant. The fit is able to
effectively constrain this conservative uncertainty due to the large
contribution from \DYtt and \ttbar in each of the categories.  At very high
masses, the uncertainty in the identification efficiency for high-\pt
\tauhadvis candidates becomes dominant, and due to the lack of significant
\DYtt and \ttbar at high mass, this uncertainty remains relatively
unconstrained. The addition of the CR-T region distributions to the fit allows
the uncertainties in the \ttbar modelling to be well constrained and as such,
they have little impact on the sensitivity.

\subsection{MSSM interpretations}

\begin{figure}[tp]
  \centering
  \subfigure[\mhmod scenario]{\includegraphics[width=0.450\textwidth]{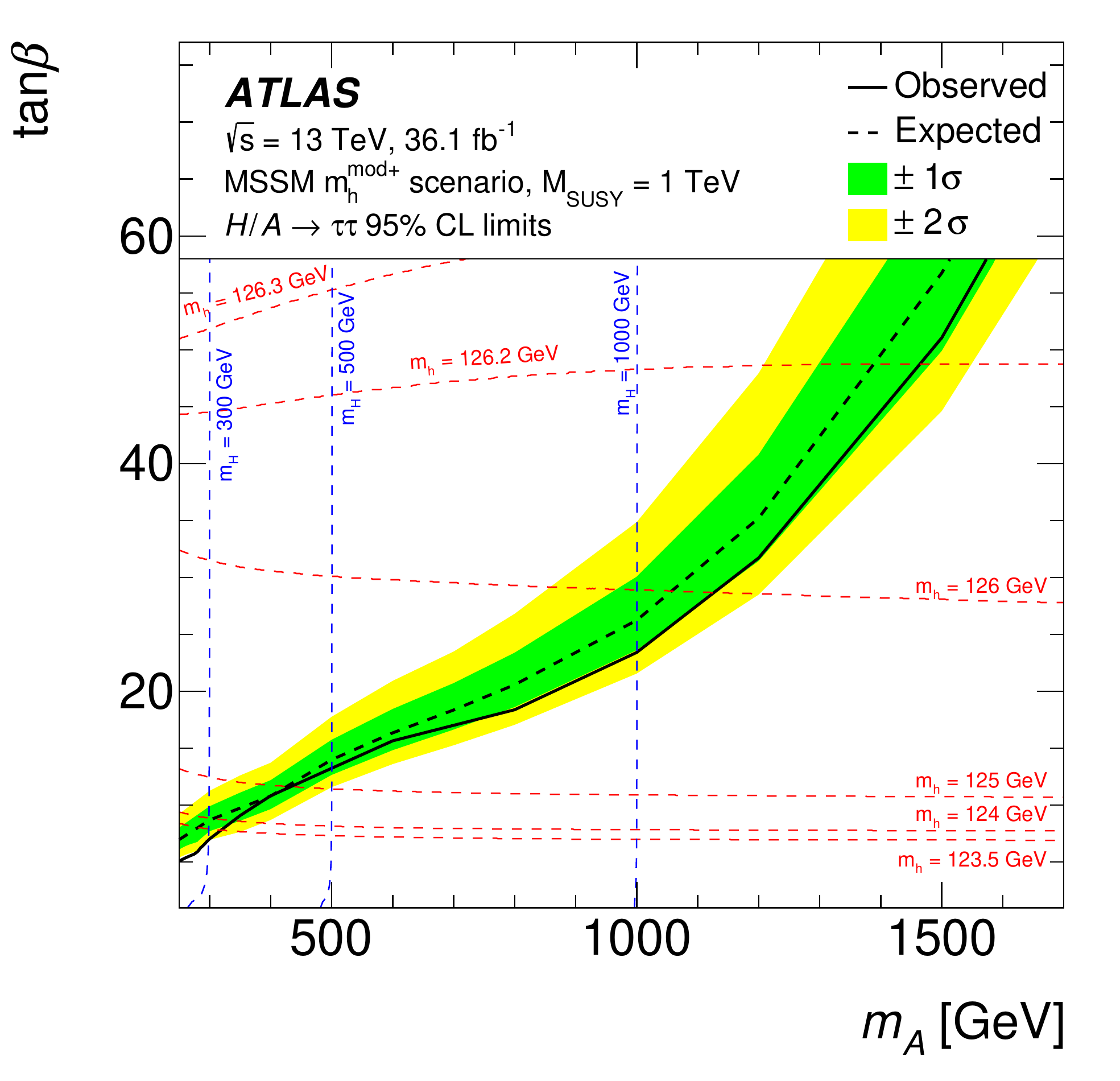}}
  \subfigure[\hmssm scenario]{\includegraphics[width=0.450\textwidth]{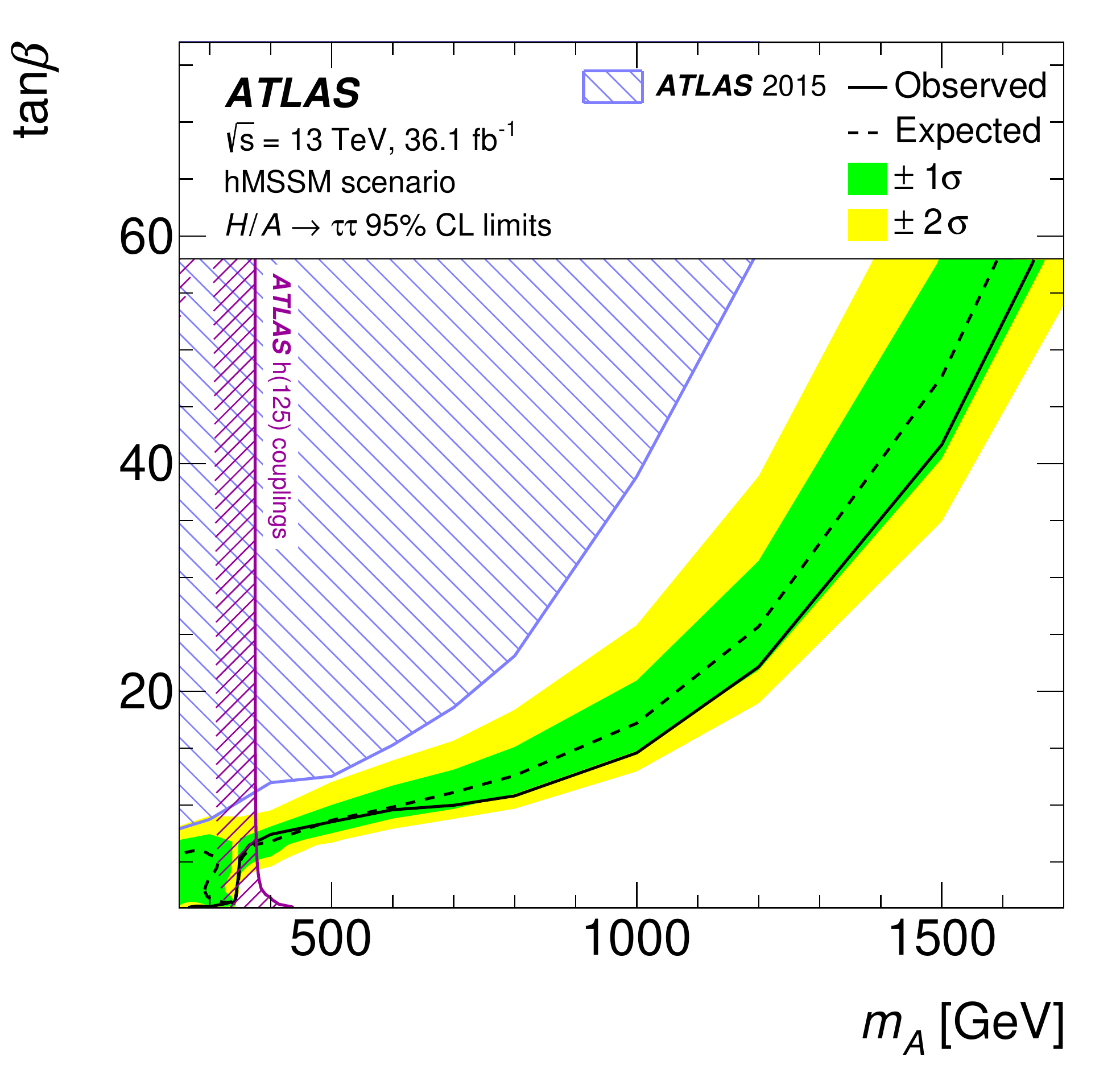}}  \\
  
  \caption{The observed and expected 95\% CL upper limits on \tanb as 
    a function of $\MA$ in the MSSM (a) \mhmod and (b) \hmssm scenarios.
    For the \mhmod scenario, dashed lines of constant $\Mh$ and $\MH$ are shown in red and blue, respectively.
    For the \hmssm scenario, the exclusion arising
    from the SM Higgs boson coupling measurements of Ref.~\cite{Aad:2015pla}
    and the exclusion limit from the ATLAS 2015
    $H/A\to \tau\tau$ search result of Ref.~\cite{HIGG-2015-08}
    are shown.
  }
\label{fig:limits-mssm}
\end{figure}

The data are interpreted in terms of the MSSM.  Figure~\ref{fig:limits-mssm}
shows regions in the $\MA$--$\tanb$ plane excluded at 95\% CL in the \mhmod and
\hmssm scenarios.
In the MSSM \mhmod scenario, the observed (expected) 95\% CL
upper limits exclude $\tanb > \limmhmodpobslow$ (\limmhmodpexplow) for $\MA =
\SI{0.25}{\TeV}$ and $\tanb> \limmhmodpobshigh$ (\limmhmodpexphigh) for $\MA =
\SI{1.5}{\TeV}$.
Constraints in the \hmssm scenario are stronger due to the presence of low-mass neutralinos in the \mhmod scenario
that reduce the $H/A\to\tau\tau$ branching fraction and which are absent in the
\hmssm scenario. 
 In the \hmssm scenario, the most stringent observed
(expected) constraints on $\tanb$ for the combined search exclude $\tanb >
\limmssmobslow$ (\limmssmexplow) for $\MA = \SI{0.25}{\TeV}$ and $\tanb>
\limmssmobshigh$ (\limmssmexphigh) for $\MA = \SI{1.5}{\TeV}$ at 95\% CL.  The
expected exclusion limit and bands around $\MA = \SI{350}{\GeV}$ reflect the
behaviour of the  $A\to\tau\tau$ branching fraction close to the  $A\to\ttbar$
kinematic threshold for low $\tanb$, allowing for some exclusion in this
region.  However, when \MA is above the $A\to\ttbar$ production threshold, this
additional decay mode reduces the sensitivity of the $A\to\tau\tau$ search for
low $\tanb$.

\subsection{\Zprime interpretations}

\begin{figure}[tp]
  \centering
  \includegraphics[width=0.450\textwidth]{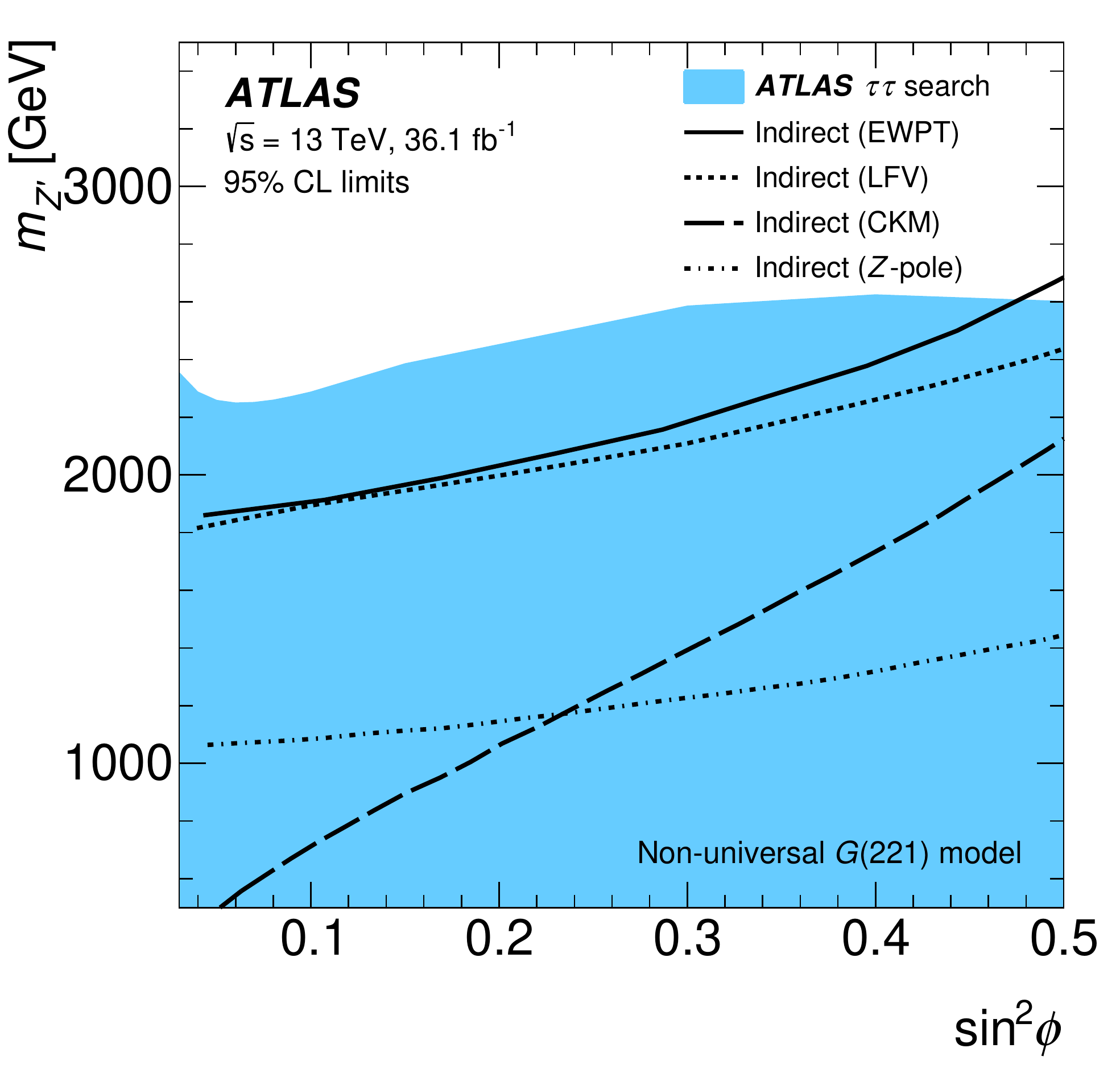}
  \caption{The 95\% CL exclusion in the non-universal $G(221)$ parameter space 
    overlaid with indirect upper limits at 95\% CL 
    from fits to electroweak precision measurements~\cite{G221LHC}, lepton 
    flavour violation~\cite{LeeLFV}, CKM
    unitarity~\cite{LeeCKM} and  $Z$-pole measurements~\cite{sfm}.  
  }
\label{fig:limits-sfm}
\end{figure}

The data are also interpreted in terms of \Zprime models.   As shown in
Figure~\ref{fig:limits-xsec3}, the observed (expected) lower limit on the mass
of a \ZPSSM{} boson is \limssm\,(\limssmexp)\,\TeV{} at 95\% CL. Limits at 95\% CL are
also placed on \ZprimeSFM{} bosons as a function of $m_{\Zprime}$ and the
mixing angle between the heavy and light SU(2) gauge groups, $\phi$, as shown
in Figure~\ref{fig:limits-sfm}.  Masses below \limsfmlow--\SI{\limsfmhigh}{\TeV} are excluded
in the range $0.03 < \sinsquarephi < 0.5$ assuming no $\mu$--$\tau$ mixing.

\FloatBarrier
\section{Conclusion}
\label{sec:conclusion}

A search for neutral Higgs bosons as predicted in the Minimal Supersymmetric
Standard Model and \Zprime bosons decaying to a pair of $\tau$-leptons is
performed using a data sample from proton--proton collisions at $\sqrt{s} =
13$~\TeV\ recorded by the ATLAS detector at the LHC, corresponding to an
integrated luminosity of \lumi.  The \ehad, \muhad and \hadhad channels are
analysed and no indication of an excess over the expected SM background is
found. Upper limits on the cross section for the production of scalar and
\Zprime bosons times the branching fraction to ditau final states are set at
95\% CL, significantly increasing the sensitivity and the explored mass range
compared to previous searches.  They are in the range
\limgghigh--\SI{\limgglow}{\pb} (\limbbhigh--\SI{\limbblow}{\pb}) for
gluon--gluon fusion ($b$-associated) production of scalar bosons with masses of
\hmin--\SI{\hmax}{\TeV} and \limzphigh--\SI{\limzplow}{\pb} for Drell--Yan
production of \Zprime bosons with masses of \zmin--\SI{\zmax}{\TeV}.  In the
context of the hMSSM scenario, the most stringent limits for the combined
search exclude $\tanb > \limmssmobslow$ for $\MA = \SI{0.25}{\TeV}$ and $\tanb>
\limmssmobshigh$ for $\MA = \SI{1.5}{\TeV}$ at 95\% CL. In the context of the
Sequential Standard Model, \ZprimeSSM bosons with masses less than
\SI{\limssm}{\TeV} are excluded at 95\% CL, while $m_{\ZprimeSFM}<
\limsfmlow$--\SI{\limsfmhigh}{\TeV} is excluded in the range $0.03 <
\sinsquarephi < 0.5$ in the non-universal $G(221)$ model.

\section*{Acknowledgements}
 

We thank CERN for the very successful operation of the LHC, as well as the
support staff from our institutions without whom ATLAS could not be
operated efficiently.

We acknowledge the support of ANPCyT, Argentina; YerPhI, Armenia; ARC, Australia; BMWFW and FWF, Austria; ANAS, Azerbaijan; SSTC, Belarus; CNPq and FAPESP, Brazil; NSERC, NRC and CFI, Canada; CERN; CONICYT, Chile; CAS, MOST and NSFC, China; COLCIENCIAS, Colombia; MSMT CR, MPO CR and VSC CR, Czech Republic; DNRF and DNSRC, Denmark; IN2P3-CNRS, CEA-DRF/IRFU, France; SRNSF, Georgia; BMBF, HGF, and MPG, Germany; GSRT, Greece; RGC, Hong Kong SAR, China; ISF, I-CORE and Benoziyo Center, Israel; INFN, Italy; MEXT and JSPS, Japan; CNRST, Morocco; NWO, Netherlands; RCN, Norway; MNiSW and NCN, Poland; FCT, Portugal; MNE/IFA, Romania; MES of Russia and NRC KI, Russian Federation; JINR; MESTD, Serbia; MSSR, Slovakia; ARRS and MIZ\v{S}, Slovenia; DST/NRF, South Africa; MINECO, Spain; SRC and Wallenberg Foundation, Sweden; SERI, SNSF and Cantons of Bern and Geneva, Switzerland; MOST, Taiwan; TAEK, Turkey; STFC, United Kingdom; DOE and NSF, United States of America. In addition, individual groups and members have received support from BCKDF, the Canada Council, CANARIE, CRC, Compute Canada, FQRNT, and the Ontario Innovation Trust, Canada; EPLANET, ERC, ERDF, FP7, Horizon 2020 and Marie Sk{\l}odowska-Curie Actions, European Union; Investissements d'Avenir Labex and Idex, ANR, R{\'e}gion Auvergne and Fondation Partager le Savoir, France; DFG and AvH Foundation, Germany; Herakleitos, Thales and Aristeia programmes co-financed by EU-ESF and the Greek NSRF; BSF, GIF and Minerva, Israel; BRF, Norway; CERCA Programme Generalitat de Catalunya, Generalitat Valenciana, Spain; the Royal Society and Leverhulme Trust, United Kingdom.

The crucial computing support from all WLCG partners is acknowledged gratefully, in particular from CERN, the ATLAS Tier-1 facilities at TRIUMF (Canada), NDGF (Denmark, Norway, Sweden), CC-IN2P3 (France), KIT/GridKA (Germany), INFN-CNAF (Italy), NL-T1 (Netherlands), PIC (Spain), ASGC (Taiwan), RAL (UK) and BNL (USA), the Tier-2 facilities worldwide and large non-WLCG resource providers. Major contributors of computing resources are listed in Ref.~\cite{ATL-GEN-PUB-2016-002}.


%

\printbibliography


\newpage
\begin{flushleft}
{\Large The ATLAS Collaboration}

\bigskip

M.~Aaboud$^\textrm{\scriptsize 137d}$,
G.~Aad$^\textrm{\scriptsize 88}$,
B.~Abbott$^\textrm{\scriptsize 115}$,
O.~Abdinov$^\textrm{\scriptsize 12}$$^{,*}$,
B.~Abeloos$^\textrm{\scriptsize 119}$,
S.H.~Abidi$^\textrm{\scriptsize 161}$,
O.S.~AbouZeid$^\textrm{\scriptsize 139}$,
N.L.~Abraham$^\textrm{\scriptsize 151}$,
H.~Abramowicz$^\textrm{\scriptsize 155}$,
H.~Abreu$^\textrm{\scriptsize 154}$
% ...
\bigskip
\\
% ...
$^{12}$ Institute of Physics, Azerbaijan Academy of Sciences, Baku, Azerbaijan\\
% ...
$^{88}$ CPPM, Aix-Marseille Universit{\'e} and CNRS/IN2P3, Marseille, France\\
% ...
$^{115}$ Homer L. Dodge Department of Physics and Astronomy, University of Oklahoma, Norman OK, United States of America\\
% ...
$^{119}$ LAL, Univ. Paris-Sud, CNRS/IN2P3, Universit{\'e} Paris-Saclay, Orsay, France\\
% ...
$^{137}$ $^{(a)}$ Facult{\'e} des Sciences Ain Chock, R{\'e}seau Universitaire de Physique des Hautes Energies - Universit{\'e} Hassan II, Casablanca; $^{(b)}$ Centre National de l'Energie des Sciences Techniques Nucleaires, Rabat; $^{(c)}$ Facult{\'e} des Sciences Semlalia, Universit{\'e} Cadi Ayyad, LPHEA-Marrakech; $^{(d)}$ Facult{\'e} des Sciences, Universit{\'e} Mohamed Premier and LPTPM, Oujda; $^{(e)}$ Facult{\'e} des sciences, Universit{\'e} Mohammed V, Rabat, Morocco\\
% ...
$^{139}$ Santa Cruz Institute for Particle Physics, University of California Santa Cruz, Santa Cruz CA, United States of America\\
% ...
$^{151}$ Department of Physics and Astronomy, University of Sussex, Brighton, United Kingdom\\
% ...
$^{154}$ Department of Physics, Technion: Israel Institute of Technology, Haifa, Israel\\
$^{155}$ Raymond and Beverly Sackler School of Physics and Astronomy, Tel Aviv University, Tel Aviv, Israel\\
% ...
$^{161}$ Department of Physics, University of Toronto, Toronto ON, Canada\\
% ...
$^{a}$ Also at Department of Physics, King's College London, London, United Kingdom\\
$^{b}$ Also at Institute of Physics, Azerbaijan Academy of Sciences, Baku, Azerbaijan\\
$^{c}$ Also at Novosibirsk State University, Novosibirsk, Russia\\
$^{d}$ Also at TRIUMF, Vancouver BC, Canada\\
$^{e}$ Also at Department of Physics {\&} Astronomy, University of Louisville, Louisville, KY, United States of America\\
% ...
$^{*}$ Deceased
\end{flushleft}



%

\end{document}